# Forensics and security issues in the Internet of Things

Shams Forruque Ahmed[a,b,*], Shanjana Shuravi[c], Afsana Bhuyian[c], Shaila Afrin[c], Aanushka Mehjabin[d], Sweety Angela Kuldeep[c], Md. Sakib Bin Alam[e], Amir H. Gandomi[f,g,*]

[a] School of Mathematical Sciences, Sunway University, Bandar Sunway, Petaling Jaya 47500, Selangor Darul Ehsan, Malaysia
[b] Department of Mathematics & Physics, North South University, Dhaka 1229, Bangladesh
[c] Science and Math Program, Asian University for Women, Chattogram 4000, Bangladesh
[d] School of Biological Sciences, Georgia Institute of Technology, Atlanta, GA 30332, USA
[e] Department of Information Technology, University of Information Technology and Sciences, Dhaka 1212, Bangladesh
[f] Faculty of Engineering & Information Technology, University of Technology Sydney, NSW, 2007, Australia
[g] University Research and Innovation Center (EKIK), Óbuda University, 1034 Budapest, Hungary

[*]Corresponding authors: shams.forruque@northsouth.edu, shams.f.ahmed@gmail.com (Shams Forruque Ahmed); gandomi@uts.edu.au (Amir H. Gandomi)

**Abstract**

Given the exponential expansion of the internet, the possibilities of security attacks and cybercrimes have increased accordingly. However, poorly implemented security mechanisms in the Internet of Things (IoT) devices make them susceptible to cyberattacks, which can directly affect users. IoT forensics is thus needed to investigate and mitigate such attacks. While many works have examined IoT applications and challenges, only a few have focused on both the forensic and security issues in IoT. Therefore, this paper reviews forensic and security issues associated with IoT in different fields. Prospects and challenges in IoT research and development are also highlighted. As the literature demonstrates, most IoT devices are vulnerable to attacks due to a lack of standardized security measures. Unauthorized users could get access, compromise data, and even benefit from control of critical infrastructure. To fulfill the security-conscious needs of consumers, IoT can be used to develop a smart home system by designing the security-conscious needs of consumers; IoT can be used to create a smart home system by designing an IoT can be used to develop a smart home system by designing a FLIP-based system that is highly scalable and adaptable. A blockchain-based authentication mechanism with a multi-chain structure can provide additional security protection between different trust domains. Deep learning can be utilized to develop a network forensics framework with a high-performing system for detecting and tracking cyberattack incidents. Moreover, researchers should consider limiting the amount of data created and delivered when using big data to develop IoT-based smart systems. The findings of this review will stimulate academics to seek potential solutions for the identified issues, thereby advancing the IoT field.

**Abbreviations**

| | |
|---|---|
| Internet of Things | IoT |
| Information and Communications Technology | ICT |
| Artificial Intelligence | AI |
| Big Data Processing and Analytics | BDPA |
| Deep Neural Networks | DNNs |
| Internet of Medical Things | IoMT |
| Federated Learning | FL |
| Encryption-as-a-Service | EaaS |
| Cyber-Physical Systems | CPS |
| Genetic Algorithms | GA |
| Central Processing Unit | CPU |
| Graphics Processing Unit | GPU |
| Deep Reinforcement Learning | DRL |
| Dueling Double Deep Q-Network | D3QN |
| Machine Learning | ML |
| Lossless Compression | LC |
| Time Processing Unit | TPU |
| Multi-Agent Collaboration | MAC |
| Adversarial Perturbations | AP |
| Permanent Faults and Timing Errors | PFTE |
| Beyond Orion Compression | BOC |

# 1. Introduction

The Internet of Things (IoT) is the interconnection of computational devices and physical objects with unique identifiers, sensors, and the ability to process and share data. The Internet of Things is revolutionizing the way we interact with technology by seamlessly connecting devices, sensors, and systems to the internet, enabling them to communicate and exchange data in real time [1,2]. It is thought to be one of the most groundbreaking technological developments of recent times [3]. This technological innovation has profoundly impacted various domains, including healthcare, agriculture, transportation, manufacturing, and smart homes [4–7]. IoT has also improved many daily life activities, such as introducing the production and consumption of data and online smart services [8–10]. Through the IoT, common devices, processes, and exchanges of information can all be upgraded to higher levels of sophistication and usefulness [11]. The IoT is equipped and able to pursue new and novel purposes, such as surveying, handling, and automating activities carried out by humans, because it employs cloud computing to merge previous services employed in apps for service-based works [12]. Users also get an opportunity to make the best choices based on their desires in accessing, handling, and surveying cloud-based resources [13]. However, not all IoT systems rely on the cloud, as some work within private networks. IoT also comes with several security issues due to the placement and distribution of nodes, and the nature of data comprised and shared over the IoT network makes it a security concern. With the rapid adoption of IoT devices across various domains, the interplay between security challenges and forensic analysis has become more critical than ever. Modern IoT ecosystems often operate within diverse environments, requiring solutions that cater to their heterogeneity, scalability, and dynamic nature [14,15]. While reviewing the applications of IoT in industrial management, Mu et al. [2] discuss several security issues regarding data integrity, cyber threats, and the heterogeneity of IoT devices. Conti et al. [16] briefly reviewed some existing security threats that come with the IoT framework. The authors discussed authentication, which is the addition of different contextual IoT devices and authenticating the routes and source of data. They also described authorization and access control, specifying rights to access different resources using only authorized devices [17,18]. As reported in this paper, certain sensitive information in IoT can affect the security of user privacy and the efficiency of IoT nodes in compiling data while ensuring privacy in case of user, content, and context. IoT architecture should also address issues and challenges when IoT devices are implemented over Software Defined Networks (SDN) and cloud environments [16].

Xiao et al. [19] introduced a blockchain-based digital forensic framework to address challenges in evidence collection and responsibility determination for industrial safety accidents involving Industrial Internet of Things (IIoT) devices. The framework employs decentralized blockchain storage for remote forensic data preservation and smart contracts for efficient evidence retrieval and tracing. A token-based access control mechanism ensures the security of IIoT device nodes, while a novel batch consensus mechanism facilitates real-time evidence acquisition. Existing studies on blockchain-based evidence preservation in IoT forensics were examined by Sakshi et al. [20]. This study found that concerns about privacy and security in IoT necessitate decentralized data storage, where blockchain provides a secure and transparent platform. Blockchain enhances IoT systems by offering privacy, authentication, and trustworthiness.

Kumar et al. [21] proposed an IoT framework called Internet of Forensics (IoF) that addresses issues like distributed computing, decentralization, and transparency in IoT forensics investigations. This solution adopts a blockchain-tailored framework, delivering a transparent view of any investigative process that includes all the entities and stakeholders in a single outline. It implements a chain based on blockchain to deal with the process, including the chain of custody and evidence chain. The lattice-based cryptographic primitives reduce complexities and are programmable, adding a layer of novelty to the idea. IoF has proven to be efficient, less complex, and time-efficient utilizes memory and CPU and is sustainable in case of energy consumption. However, blockchain technology faces security and immutability issues with private keys and data, as well as privacy concerns, because all network nodes have access to data. A blockchain-based approach developed for a smart home domain was presented by Brotsis et al. [22], which deals with collecting and preserving digital forensic confirmation. Forensic evidence is maintained in an encrypted private database, and the system also uses a blockchain authorized by the court to provide security features like integrity, authentication, and non-repudiation. The blockchain holds the extensive metadata of evidence for providing the services above, and it communicates via smart contracts with various devices incorporated into an investigative process, such as Internet service providers (ISPs), law enforcement agencies, and lawyers. IoT forensic evidence processing presents unique challenges, but this framework offers a high-level architecture of blockchain-based solutions to address them. Yet, the paper did not address the same blockchain-based problems, such as privacy vulnerability and data immutability, and the proposed framework's way to tackle them.

Shruti et al. [23] explored innovative approaches to address data security and access control challenges in fog computing environments. While CP-ABE, a widely used cryptographic technique, provided precise access control, existing multi-authority CP-ABE methods were unsuitable for resource-constrained IoT devices due to the growing sizes of secret keys and ciphertext with additional attributes. To resolve this, the study proposed a novel multi-authority CP-ABE (MA-based CP-ABE) approach that optimized the lengths of secret keys and ciphertext, ensuring constant secret key size regardless of the number of attributes. A fog-based IoT forensic framework was introduced by Masri et al. [24] that works to tackle the key challenges within digital IoT forensics. The fog computing concept is used in this framework to distribute nemanyintelligence to previously inaccessible nodes. This architecture is suitable for IoT systems with many interconnected nodes and a lot of data to process. Forensic investigations relying on data connections between IoT entities and a fog node or gateway should have a better chance of recovering relevant evidence using this framework. When this framework investigates, data are analyzed, and if any suspicious activity is determined, the IoT nodes or entities are notified of potential risk. In this way, the threat does not move to the other devices. However, fog and edge computing have certain drawbacks, such as increased design complexity, decentralized design, and entities placed in less secure conditions, which were not addressed in this paper.

Despite numerous forensic frameworks and techniques proposed over the years, addressing the diversity and complexity of IoT devices and networks remains a significant challenge. Mahmood et al. [25] conducted a comparative analysis of existing forensic investigation frameworks, evaluating their strengths and weaknesses across ten key parameters, including heterogeneity, scalability, and chain of custody. The importance of IoT in the case of forensics in identifying, compiling, preserving, and reporting the evidence was reported by [16]. Given the autonomous and passive nature of IoT devices, it is hard to detect their presence due to the lack of well-reviewed methods to compile remaining evidence from devices in a forensically acceptable way [26,27]. Obtaining evidence from IoT devices is extremely difficult due to the lack of time information stored by most IoT nodes and the massive volume of data. The lack of reviewed methods, forensically acceptable tools for compiling, preserving, and processing systems data, and the lack of attribution of malicious activities present a significant challenge for IoT forensics [28]. The applications of IoT in sectors such as healthcare, education, homes, power distribution, cloud, web and mobile applications, agriculture, and technology were explored in various studies [29–32]. The challenges with IoT applications have also been explored, including security, privacy, heterogeneity, lack of standardized protocols, device identity, vulnerability detection, and scalability. Rehman et al. [33] identified the research challenges faced with IoT. These include the need for universal standards to deal with the heterogeneous climate of the devices and the challenges that come with their lack thereof, technical challenges with the network, architecture, and cellular computing, and the problems arising with data privacy.

To investigate the forensic and security issues related to IoT implementations across different disciplines, an integrative literature technique was conducted in this study. This study aims to fill the existing gaps in the literature by providing a comprehensive review of the forensic and security issues in IoT systems. By systematically analyzing various frameworks, challenges, and technological advancements, the study highlights the areas requiring immediate attention from researchers and practitioners. Specifically, relevant and reliable documents were gathered, sorted, inspected, and analyzed. We conducted our database searches using only the most prestigious resources available, including Scopus, Google Scholar, and the journals published by Elsevier, Taylor & Francis, Wiley, Sage, Springer, ACS, Frontiers, MDPI, and Inderscience. Relevant literature was identified using various keywords, including Internet of Things, IoT, Security Issues, Forensic Issues, IoT Architecture, Cyber Security, and IoT Applications. While many reviews have covered IoT applications, research challenges in IoT, IoT forensics, or security issues, there is a lack of a comprehensive review that incorporates IoT architecture, forensics, security issues, research challenges, and the future of IoT, as illustrated in Table 1. In the present context of the 21$^{st}$ century, this article seeks to combine the applications and security issues along with forensics challenges and prospects of IoT. This review addresses the fundamental architecture of IoT, the forensics and security issues associated with IoT, the numerous uses of IoT, and the research challenges and prospects of IoT. It provides a comprehensive compilation of the most pertinent IoT-related literature and a discussion of the many distinct facets of IoT, with sufficient correlations to establish a bridge between the existing body of knowledge.

Table 1. Comparison of the current review, which focuses on forensic and security issues in IoT, with other relevant reviews from 2018–2024.

| Review study | IoT architecture | IoT Forensics | IoT security | Research challenges in IoT | Future prospects toward IoT |
|---|---|---|---|---|---|
| This study | √ | √ | √ | √ | √ |
| Conti et al. [16] | × | √ | √ | × | × |
| Stoyanova et al. [34] | × | √ | × | × | √ |
| Atlam et al. [35] | × | √ | √ | √ | √ |
| Yaqoob et al. [36] | × | √ | × | √ | × |
| Mahmoud and Aouag [37] | × | × | √ | × | √ |
| Ogonji et al. [38] | × | √ | √ | √ | × |
| Tahsien et al. [39] | × | × | √ | √ | × |
| Aly et al. [40] | × | × | √ | √ | × |
| Azmoodeh et al. [41] | × | √ | √ | √ | √ |
| Yang et al. [42] | × | √ | √ | √ | × |
| Mahmood et al. [25] | × | √ | √ | × | √ |
| Qureshi et al. [43] | √ | √ | √ | × | √ |

√: available; ×: unavailable

## 2. Internet of Things architecture

The Internet of Things (IoT) is a network of interconnected devices/objects/mechanical and digital machines/etc.—"things"—comprising sensors, software, and other such technologies that exchange data [44,45]. In its most basic form, an IoT architecture is a collection of components that includes sensors, protocols, controllers, cloud services, and layers. In addition to devices and sensors, the IoT architectural layers are separated by protocols and gateways that track the continuity of a system [46,47]. There is no widely accepted IoT architecture; numerous architectural alternatives have been proposed. A thorough overview of the present state-of-the-art IoT architectures across different domains can be found in [11]. The impact of IoT on people's lives today is unprecedented as it is implemented in healthcare systems, wellness, education, amusement, social activities, energy efficiency, environmental sensing, remote monitoring, and transportation systems, among many other industries.

Nevertheless, an architecture using IoT and other technologies is needed to improve the resilience of the IoT infrastructure itself. A novel IoT architecture was suggested by Abreu et al. [48] suggested a novel IoT infrastructure, including IoT middleware, services, and architecture layers, to satisfy the criteria while ensuring the quality and robustness of the services and infrastructure. Routing Protocol was implemented for Lossy networks, and mechanisms were incorporated to wrap the IoT infrastructure details (Heterogeneity Manager, Communication Manager, Virtualize Device Manager, IoT Service Manager) [48]. The model utilizes the cloudlet and cloud approaches to increase the environment's proliferation and extensibility. Additionally, the architecture demonstrated that resilience may be boosted by merging the proposed structure with the appropriate processes and protocols. Consequently, many diverse situations can be improved in a smart city to benefit inhabitants, such as traffic control, emergency health support, and so on [48]. However, one of the issues that can emerge is the maintenance of the services' robustness, which must be agile and flexible to scale. Security and privacy are the key concerns since most connections with objects are vulnerable to hacks.

Chen et al. [44] proposed model checking and timed automaton modeling for the IoT architecture and introduced the theory and process of modeling with time automata. The authors verified the IoT system in the older population's health cabin using the UPPAAL validator. The results addressed deadlock system activity, and the design of the elderly health cabin did not waste any resources. The authors took into account the physical conditions of both summer and winter. The system problem was efficiently detected and signaled by the cloud control services. The model, however, did not create a multidimensional representation of the IoT system, nor was the modeling formalism, one of UPPAAL's worst flaws, tackled.

IoT environments generate enormous amounts of data, which raises security vulnerabilities, including sensitive data and network manageability. Javaid and Sikdar [49] propounded a blockchain framework that uses a blocking checkpoint technique and dynamic proof of work consensus to address this. Unlike standard Proof of Work (PoW) based mining methods, which often use constant difficulty levels, the Dynamic PoW (dPoW) functioned with fluctuating difficulties in data extraction. As a result, the design was able to efficiently take into account increases in data transmissions in IoT contexts and the related hardware. Moreover, enhanced security and a shorter block mining time are two benefits of using checkpoints instead of traditional PoW to mine a block. However, the model did not shed extensive light on the fact that with evolved technology, many new security problems could influence the protocol layers.

**2.1 Three- and five-layer architectures**

During the early phases of IoT research, the most fundamental architecture in this field was developed. It contains three layers: application, network, and perception [50], as shown in Fig.1. The perception layer is made up of physical components that have sensors for observing and collecting information about the environment. It finds specific physical characteristics and recognizes other states of technology nearby. The network layer manages connections to services, networking hardware, and other electronic objects. Furthermore, its skills are utilized for distributing and analyzing sensor data. The task of offering the users software-specific services falls under the purview of the application layer, which describes a variety of IoT applications. However, this architecture is inadequate, given the growing breadth of IoT.

Bary et al. introduced a Multi-Layer (four-layer) Blockchain Security Model designed to enhance the security of IoT networks. The model not only safeguards IoT networks but also aids in their implementation to protect similar networks, effectively preventing transaction privacy leakage for all users within a public blockchain network. The model leverages the clustering concept to support its multi-layer architecture[51]. The smart hybrid learning method (SHLM) is a learning approach that adapts to the Education 4.0 system and allows for stakeholder involvement. Hartono et al. [52] developed a model and architectural framework design for SHLM using a three-layer architecture approach. Two parts of the framework were the suggested model based on the three-layer architecture and the conceptual framework. According to the conceptual framework, SHLM is a learning network portal that links educators, pupils, professionals, businesses, localities, and the government. Through the system, the students could navigate cases and ongoing projects and interact with businesses, professionals, and communities using smart conferences, which were available on the portal. Along with the design evaluation report, the finished product must be uploaded to the system. The system is divided into three layers: View layer, which provides user interface displays and is accessible from anywhere at any time via a browser; Domain layer, which offers features like material access, scheduling, threads of discussion, videos, and learning groups; and Data Access layer, which provides data access (allowing users to backup, read, update, create, restore, and connect data to a database object). Students are able to learn outside of the classroom, thanks to SHLM. However, the authors did not consider the growth of each student's unique professional path at their time of use, as the future guidance of each student is not the main focus of the model.

A deep learning-based framework was developed by Kethineni & Pradeepini [53] to detect intrusions in smart farming systems. The architecture was three-tiered, with the first tier being the sensor layer, involving the placement of sensors in agricultural areas. The second tier was the Fog Computing Layer (FCL), which consisted of Fog nodes where the proposed IDS was implemented. The gathered information was transferred to this fog layer for further data analysis. The third tier was the cloud computing layer, which provided data storage and end-to-end services. The proposed model included a fused CNN model combined with a bidirectional gated recurrent unit (Bi-GRU) model to detect and classify intrusions. As the demand for more complex, higher-quality products increased, the manufacturing sector became more automated and computerized. Parto et al. [54] introduced a ground-breaking design for a scalable IoT that allows Industry 4.0 to thrive more effectively. The three layers of cloud, fog, and edge were presented as part of an innovative IoT architecture for Cyber-Physical Systems (CPS). Operator inputs, embedded systems, and other sources of data from industrial applications were found at the edge layer. The fog layer's primary responsibilities are cloud and local communication, stream analytics, and incremental machine learning. At the cloud level, federated learning and communication are the two key goals of the cloud layer. In the mentioned study, the mean and standard deviation of datasets were calculated using aggregation procedures. A Bayesian paradigm for vibration analysis was presented as an example application to demonstrate the application of the suggested framework. The outcomes demonstrate that the trained models from the sites might be successfully aggregated without sending the raw information to the cloud. The development of computational IoT frameworks with better expandability and privacy using the suggested architecture was successfully shown using the vibration recognition framework, Federated Learning algorithms, and Bayesian classification [54]. However, the model falls short in terms of scalability to meet the ever-increasing demands. Also, the study failed to mention continued efforts to keep the architecture up-to-date by enhancing performance efficiency and reducing the storage, load, and bandwidth necessities for the cloud.

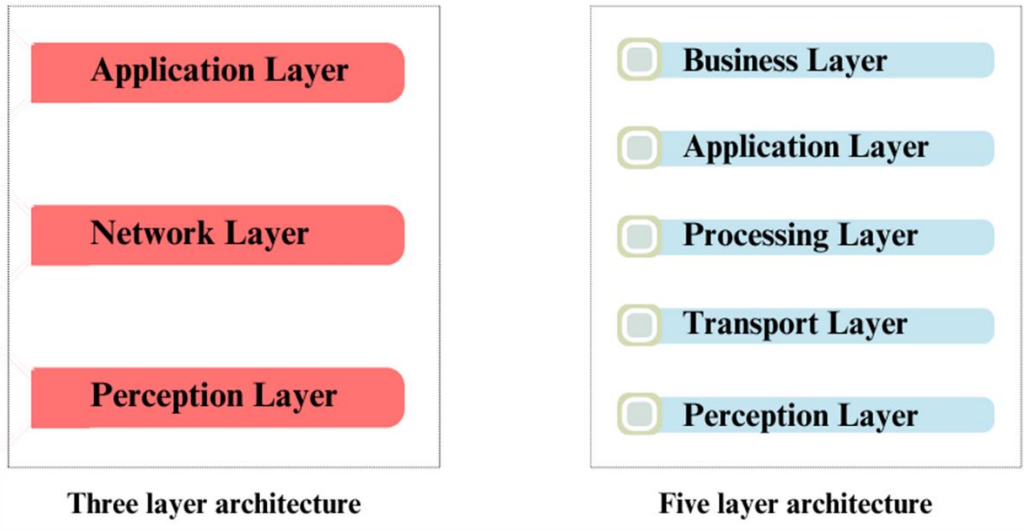

Fig. 1. Three- and five-layer architectures

The five-layer architecture comprises perception, processing, transport, business, and application layers [55]. The perception and application levels serve the same purpose as the three-layer layout, and the responsibility of the other three layers is subsequently described. The transport layer uses networks to move visual information from the perception layer to the processing layer and back. The middleware layer also referred to as the processing layer, collects, analyzes, and evaluates high volumes of transport layer data. It has the ability to oversee and offer a variety of facilities to the different layers. The business layer oversees the IoT infrastructure, including applications, services, business models, and privacy protection. More specifically, the layers of this architecture include the physical perception, protocol and network, transport, application, and cloud and data services layers. In one study, a Wireless Body Area Network (WBAN), Wireless Sensors Network (WSN), Radio-Frequency Identification (RFID), and other sensors and IoT devices were used in the physical layer of the suggested compact design [50]. An IoT system's network and protocol layer includes ZigBee, Ethernet, Wi-Fi, Bluetooth, Ethernet, and 5G, as well as other network communication protocols. The transport layer is made up of IP plus the suite of protocols known as Transport Layer Security (TLS) or Secure Sockets Layer (SSL) [50].

Zhu et al. [55] suggested a five-layer architecture intending to create a de-facto framework to turn massive data into useful information. To develop modern Big Data Processing and Analytics (BDPA) systems, a vast number of static data and online data streams were collected and analyzed. The study highlights the most recent technology and solutions while illuminating the five layers' difficulties and functionalities. The system requires a fundamental, scalable storage infrastructure for storing historical data as well as moving data from the online data stream. An analytics layer is necessary to gather information, forecast, and user-tailored activities. According to their research, an application layer must act as a configurable layer for the entire system so that the architecture can be adaptable enough to meet the needs of various industries. The system accepts both batch and streaming data inputs. The output of the architecture is generated at the analytics layer, where the results can be used for attribute testing and data retrieval [55]. Instead of addressing specific issues, the design addresses the analytics issue. However, less user-friendly BDPA solutions are now a drawback of this approach, in addition to the challenges associated with privacy, labelling, fault tolerance, hot-plugging, and generalizability.

**2.2 Cloud and fog-based architectures**

In order to analyze and manage the massive amounts of data created by connected devices, cloud and fog-based architectures have emerged as two important concepts in the IoT field [56,57]. Data collected by IoT devices is sent to remote cloud servers where it can be stored, processed, and analyzed [58]. Large-scale deployments of the IoT can be managed effectively by cloud platforms due to their scalability, storage capacity, and computational power. They facilitate cutting-edge analytics, artificial intelligence (AI)/machine learning (ML) algorithms, and data visualization, ultimately allowing companies to gain useful knowledge from IoT data. When dealing with real-time and time-sensitive applications, however, cloud-based architectures might run into problems, including latency, network congestion, and higher bandwidth needs [59,60].

As a supplementary method for resolving these issues, fog-based architectures have become prominent. Fog computing relocates data processing, storage, and communication closer to the network's periphery, which is physically closer to the data itself [61]. Data processing occurs closer to the data sources, decreasing latency and bandwidth usage. To facilitate rapid responses and decisions, fog nodes, located at the network's periphery, are capable of real-time data analysis, filtration, and pre-processing [62,63]. Autonomous vehicles, smart city infrastructure, and industrial automation are some of the applications that might benefit greatly from fog architectures' low-latency interactions.

System architectures for fog and cloud computing are somewhat different in terms of data processing and security. Cloud computers in various system topologies process data in a large, centralized manner, where the cloud is in the center with applications stacked upon it and a network of intelligent objects underneath, as illustrated in Fig. 2. Due to its high level of scalability and flexibility, it is given priority [46]. However, network interfaces, detectors, and sensors handle several data acquisition and analyses in fog computing. In a fog architecture, the pre-processing, monitoring storage, and security layers are positioned between the transport and physical levels [64].

Hybrid architectures that combine cloud and fog computing allow organizations to utilize the benefits of both approaches [65]. While fog nodes can process data at the edge for rapid actions, more resource-intensive tasks, such as in-depth analysis and long-term storage, can be offloaded to the cloud. This combined methodology increases the scalability and effectiveness of IoT systems by making better use of available resources and decreasing reaction times.

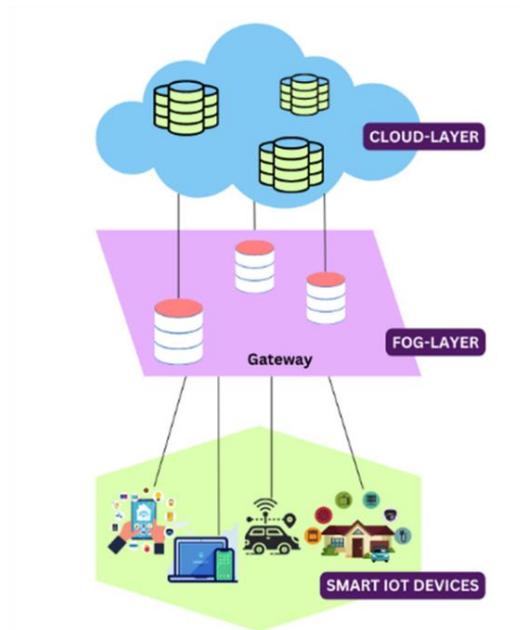

Fig. 2. Cloud and fog-based architectures

As the volume of sensitive data increases, IoT has increased demand for computation, networking, and storage capacities. However, this comes at a cost because unnecessary connections strain both the cloud data centre and core network. To address the aforementioned difficulties, Razouk et al. [46] suggested a unique middleware architectural model. They described the general notion of combining fog computing and cloud computing to reach greater security. The model contains a Security Module comprising access control, data/privacy protection, authentication, and security profiles. Next, the Communication Module comprises a decision-making module, an API module, and IoT applications with cloud and fog computing [46]. The security features of the proposed framework include the implementation of data integrity checking, forward secrecy, replay attacks protection, impersonation attack protection, traceability protection, mutual authentication, and so on. However, the study does not mention the extensive application of this framework, be it in test beds and real-world scenarios. Another concern of the model is the constant update required to improve performance, efficiency, and reliability. The security architecture also needs to be up-to-date as there is an increase in various advanced attacks.

Current Encryption as a Service (EaaS) architectures fails to effectively leverage cloud and fog layers, limiting their potential for performance enhancement. To address this, Javadpour et al. [66] introduced a novel EaaS architecture, termed full-cloud-fog, designed to boost EaaS throughput. This is achieved by positioning frequently accessed components in the fog layer while utilizing cloud nodes to manage resource allocations. The security features of the proposed architecture were thoroughly analyzed, followed by implementation on a real-world testbed. Evaluation results demonstrate that the full-cloud-fog architecture increases EaaS throughput by 81%. An offloading technique to enhance the efficiency of IoT applications within a three-layer architecture in a fog computing environment was proposed by Abdulazeez & Askar [67]. Due to the limited storage and computational capabilities of IoT devices, task offloading to other layers was essential to ensure efficient processing and meet Quality of Service (QoS) requirements. To address this, a fuzzy logic-based task scheduler was employed to make informed offloading decisions by evaluating task attributes and determining the appropriate processing layer—locally, at collaborative fog nodes, or in the cloud. A Deep Q Network (DQN) method was also utilized to optimize fog node selection and maintain balanced workload distribution among collaborative fog nodes.

Li et al. [68] developed a framework for cloud-fog-local video encryption as part of their research. It features an extensive video encryption framework based on Spark, a three-layer service based on management approaches, and a fine-grain video encryption technology based on the Network Abstract Layer Unit (NALU). A "cloud-edge-local" media cloud service hierarchy was introduced by the authors. As the brain of the entire system, the cloud layer serves two purposes: i) sending visual processing instructions to the fog layer was the initial stage, and ii) authenticating devices and verifying their legitimacy at the local layer. Before encrypting and decrypting the videos, the fog layer disperses computer power for adjacent devices. The local layer represents multiple user terminals, transmitting video encryption and decryption requests to the cloud layers. The three-level architecture of the suggested model was matched by the authors' hierarchy key management approach. The research assessed the proposed fine-grained video encryption system's encryption effectiveness, security, and compression loss. The testing and analysis conducted in the study proved that the suggested video encryption approach complies with security protocols regarding private and public videos. However, it did not result in any deformation or compression loss. Drawbacks do exist, though, as, in this scenario, video content was more informative than written content. As a result, additional issues concerning copyrights, intellectual, and data leakage may arise. Another challenge concerns securing it against multiple attacks at once.

To deal with the delays in cloud-based technologies, Gupta et al. [64] developed a model based on fog-cloud architecture for hospitals to reduce energy usage, latency, and delay. The proposed method has a three-layer design, including healthcare IoT sensors, fog, and cloud layers. Specifically, the first layer of healthcare IoT sensors uses location, ultrasound, and other sensors to gather information about a patient's health. After that, the cloud and IoT sensor layers for healthcare are positioned in the middle. Third, the data are kept in a cloud data centre by the cloud layer. As per the application, the outcomes based on this model showed a reduction in the death rate. The model does not, however, include a security mechanism that might improve the architecture's speed and performance. There is also seemingly a need for an update to reduce delay.

The network produces more data as distributed energy resources (DERs) and IoT devices permeate distribution networks (DNs). Yue et al. [69] presented a state-of-the-art energy-efficient approach, the cloud-fog hierarchical architecture, in response to the rising power management requirements of the upcoming generation distribution networks. By receiving raw data from the prosumers and basic users' units in the DN, the fog computing layers perform regression and grouping analysis predictions. The cloud computing layer is used to achieve the overall objective of optimization. The system's hierarchical cloud and fog layers work together to control energy while prosumers and customers wait at the DN's terminal. The authors used the Genetic Algorithm (GA), a powerful tool for large-capacity nonlinear and discrete problems, to tackle the global optimization problem posed by the cloud employing cloud and fog layers and terminal units. GA simulates biological reproduction (survival of the fittest) through mutations, interchange, and natural selection by coding all possible solutions into a vector, each element of which is referred to as a gene, then evaluates each gene in the group using the objective function. The model demonstrated that the energy management architecture makes it possible to arrange and manage various operating information in the DN in real time [69]. The technology will enable autonomous energy management and hourly outcomes for the following DN over the years. The model, however, lacked diverse stakeholders, cutting-edge artificial intelligence, optimal decision-making, and intelligent energy management. The study did not investigate the model's performance under unpredictable network conditions and heterogeneous communication technologies. Table 2 summarizes the three and five-layer architectures & fog and cloud-based architecture.

Table 2. Summary of the three and five-layer architectures & fog and cloud-based architecture.

| Study Ref. | Architecture Type | Key Features | Remarks |
|---|---|---|---|
| [51] | Multi-layer Blockchain | Enhances IoT network security and prevents transaction privacy leakage using clustering concepts. | The model leverages the clustering concept to support its multi-layer architecture. |
| [52] | SHLM Three-layer | Enables stakeholder collaboration in Education 4.0; View, Domain, and Data Access layers support outside-classroom learning and interaction. | Neglects individual student career growth and professional path guidance. |
| [53] | Three-tier (Smart Farming) | Sensor, Fog Computing, and Cloud layers use fused CNN and Bi-GRU | Lacks scalability to meet increasing demands. |

| | | for intrusion detection in smart farming. | |
| --- | --- | --- | --- |
| [54] | Three layer (Cyber-Physical Systems) | Innovative edge-fog-cloud IoT architecture integrates federated learning and Bayesian analysis for real-time analytics and data privacy. | Scalability challenges; lacks updates for storage, load, and bandwidth efficiency. |
| [55] | Five-layer IoT | Integrates batch and streaming data inputs, scalable storage, and an adaptable application layer to enable modern Big Data Processing and Analytics systems, providing industry-tailored insights through a robust analytics layer. | Limited user-friendliness; challenges with privacy, fault tolerance, and generalizability. |
| [46] | Cloud-Fog Hybrid | Middleware security model with access control, authentication, and data integrity modules. | Requires constant updates; lacks extensive real-world application validation. |
| [66] | Full-Cloud-Fog EaaS | Boosts Encryption-as-a-Service (EaaS) throughput by relocating components to fog layer; real-world testbed validation. | The proposed architecture increases EaaS throughput by 81%. |

| [67] | Task Offloading in Fog within a three-layer architecture | Fuzzy logic and Deep Q Network optimize task scheduling and workload distribution in a three-layer IoT architecture. | Requires further efficiency improvements for large-scale deployments. |
| [68] | Cloud-Fog-Local Video Encryption | Uses Spark for encryption; hierarchical architecture with fine-grained control; balances workload across layers. | Vulnerable to copyright and data leakage; lacks multi-attack resistance. |
| [64] | Fog-Cloud for Healthcare | Reduces latency and energy usage in hospital IoT systems; three-layer architecture with healthcare sensors, fog, and cloud layers. | No security mechanisms; potential for optimization in performance and speed. |
| [69] | Cloud-Fog for Energy Management | Uses hierarchical architecture with Genetic Algorithms for real-time distribution network optimization and autonomous energy management. | Lacks cutting-edge AI, stakeholder diversity, and performance evaluation under unpredictable conditions. |

## 2.3 Social IoT

A new paradigm called the social IoT (SIoT) analyzes the social connections between inanimate items like humans do. One strategy to navigate the IoT is to start with one device and then move on to the other connected devices. SIoT also demands reliability and connection strength. Using techniques to study human social networks, we can look at the social networks of IoT devices. ID, meta-information, security control, service discovery, relationship management, and service composition are some of its fundamental components [70–72]. The server side of the SIoT architecture contains three layers: foundation, component, and application.

A study by Khan et al. [73] focused on developing intelligent and efficient SIoT technologies to analyze human behaviors, enhancing activity accuracy and user performance. Nine key features were evaluated to assess their potential impact on human behaviors and supporting facilities. Seven SIoT innovations and devices were explored as alternatives to improve human behaviors. The EWM method was employed to quantify the accuracy of SIoT studies, while the TOPSIS methodology served as a reliable selection approach for identifying the best SIoT solution. The findings revealed that MCDM- and SIoT-based technologies were effective tools for evaluating and advancing SIoT-driven human behavior analysis. To safeguard the 5G-IoT framework's digital solutions, Ullah et al. [74] suggested a hybrid approach for the Control Flow Graph (CFG) and a deep cognitive approach. After extracting the newly given APK files, the JDEX decompiler was used to retrieve Java source documents from potentially duplicated and genuine apps. After that, the source code was divided up into distinct Android-based modules. After creating the control flow Graphs (CFGs), the graded features were extracted from each module. In order to better predict duplicated products, the Recurrent Neural Network (RNN) was created, which works by learning features from various Android application modules. The proposed technique was compared to existing approaches like Random Forest (RF), Multi-Layer Perceptron (MLP), and Convolution Neural Network (CNN). Classification accuracy, recall precision, and F-measure were separate performance criteria used to compare each approach across four different cloning applications. For cloned programs selected from multiple Android app stores, the suggested strategy exhibited an average accuracy of 96.24% [74]. However, it is not clarified how this method will be sustained in the long run, be it Android or other mobile operating systems. While the hybrid method follows instructions in textual form, the image-based impact has not been mentioned. In case of an attack on IoT systems, the security protocols seem lax.

Using the Atlas thing architecture and the thing (IoT device description language) IoT-DDL project, Khaled and Helal [75] proposed a cross-interactions programming framework that creates a decentralized programming environment/system for SIoT. A collection of specific associations is included in the framework, which expands the social IoT thing-level links and enables the development of a far wider variety of beneficial IoT applications. The model's architecture utilizes operating system services to provide extra capabilities required for ad hoc connections and interactions. The architecture is built on top of the IoT-DDL standards, an XML-based descriptive language. The specialty of this language is that it specifies a thing's identity along with its relevant applications that may be uploaded to it. The Atlas IoT platform layer of the framework concentrates on objects' semantic and descriptive properties to increase item engagement and programmability. The three components needed for an IoT app in the suggested inter-thing system include: recipe, Thing Relationship (TR), and Thing Service (TS). The framework incorporates three methods for wiring up and merging the various primitives: match, filter, and evaluate. However, the model does not mention the approach of the Atlas framework when it comes to its overall scope, different architectural layers, IoT protocols, and homogeneous communicating things.

IoT allows any gadget to interact with a user who can be individually identified. Interactions and relationships between gadgets and their users may be rather complex. Stelea et al. [76] developed cooperative service models and approaches based on societal communication between users and services. The SIoT paradigm and related interaction mechanisms serve as the foundation for these frameworks and approaches. The SIoT architecture for dynamic service composition is comprised of four functional levels: Aggregation, Virtualization, Real World, and Application Users' macro services. The ability of the Virtual Socially Aware Objects (VSAO) was increased with increased trustworthiness and control thanks to the Socially Aware Parameters (SAP), which were utilized to produce social behaviour in virtual socially-aware objects. While this enables the abstraction of real-world services into simulated objects, their study provides limited information regarding its privacy implications. If technology simulates human behaviour, the question remains as to whether it also imitates its negative, radical, and sensitive characteristics.

Controllability is a critical issue in complex networks with significant applications in social, biological, and technical systems. Although it plays an essential role in managing social smart cities, it had not been explicitly defined as a problem within SIoT networks, nor had solutions been proposed. Aghee et al. [77] addressed the controllability problem in temporal SIoT networks. A formal definition of temporal SIoT networks was first introduced, along with their unique relationships, which were modeled systematically. The controllability problem was then applied to these networks (CSIoT) to identify the Minimum Driver Nodes Set (MDS). The proposed CSIoT approach was compared with state-of-the-art methods for performance evaluation. The results revealed that heterogeneity, including variations in types, brands, and models, was analyzed, with 69.80% of the SIoT sub-graph nodes identified as critical driver nodes across 152 sets. The proposed approach enabled distributed network control in temporal SIoT networks.

Privacy disclosure issues exist alongside cloud computing advancements. Therefore, a Persian framework was proposed by Bi et al. [78], offering individualized services to social IoT users while maintaining user privacy. Static Bayesian game theory was used in the strategic conflict between Cloud Service Providers (CSPs) and vulnerabilities to counter opponents with prior knowledge. Its main objectives are implementing user classification, outlining the trust and security obligations of CSP, and striking a decent mix between Quality of Service (QoS) and Privacy Protection [78]. The users deduce their privacy choices independently using offline fuzzy reasoning. TMC oversees the CSP's confidence to implement effective service operations. The CSP offers its users individualized attention and service based on their unique tastes. Additionally, the authors achieved privacy protection by utilizing the game-mixing strategy equilibrium. Meanwhile, using the suggested paradigm, the authors assessed privacy disclosure using information entropy [78]. On the other hand, their study did not consider further adversaries in the IoT world like incorrect access control, application vulnerability, outdated software, etc. The accuracy of fuzzy reasoning resembling human reasoning was also not examined.

## *2.3.1 Basic components*

Social IoT users have made extensive use of search engines, some of that is created specifically for SIoT. The growth of SIoT is being driven by search engines, which generate shortcuts between network nodes inside communities, including between two independent SIoT. Fu et al. [79] presented a search SIoT (SSIoT) model that took into consideration both link-adding and node-joining. A social network layer and an IoT layer are included in the proposed two-tier SIoT paradigm. The SSIoT model was built using social science theories that have undergone empirical testing, including Breiger's ground-breaking research on two-dimensional concepts of social networks. A search engine, respective users, and objects are the three components of the SSIoT paradigm. The six measures utilized to describe changes in network structure were average distance, network stability, network diameter, degree distribution, network density, and user betweenness. The study further demonstrated that the degree distribution in SSIoT follows an improved power law, and the mentioned measures demonstrated an increment. The authors then tested their models for each metric. However, the model performed poorly in real-world and practical applications. Thus, a more quantitative and accurate approach is needed regarding the relationship between SIoT structures and search engines.

### 2.3.2 Representative architecture

There are several advantages to using a clustered design. By lowering the number of records for each limited object, a clustering design can save memory usage. Abderrahim et al. [80] presented the Trust Management Community of Interest Social Internet of Technology (TMCoI-SIoT), a trust management system (TMS) for SIoT. The study considered a variety of aspects, including indirect and direct trust, social trust modelling, and transactional considerations. Each cluster in their study had a central element called the admin, which was in charge of calculating and documenting the community members' levels of trust. The suggested TMS included filter-based trust prediction as well as trust calculation at the node and admin levels. Following the computation of direct and indirect observations, an overall trust calculation was performed, and trust levels were updated. The authors ran Python simulations to demonstrate the robustness, efficacy, and dependability of TMCoI-SIOT in an IoT context [80]. Their work was comparable to numerous other current efforts in terms of efficiency. Their study, however, did not clarify how the model will recognize and survive in the face of more attacks. Its shortcomings should be emphasized so that numerous viable responses might be proposed.

In summary, each possible architecture for the IoT has its uses and benefits. The implementation of an IoT system can benefit from a structured framework provided by IoT architectures like the three- and five-layer architectures. Fog-based architectures enable edge computing, while cloud-based ones use the cloud for data storage and analysis. Social IoT architectures encourage people and things to work together and share information. The following table (Table 3) summarizes the discussed concepts, methods, and findings of Social IoT (SIoT).

Table 3. Summary of Social IoT

| Aspect | Details |
|---|---|
| Definition | SIoT is a paradigm that examines social connections between devices, akin to human social networks, to improve device interactions and reliability. |

| Fundamental Components | ID, meta-information, security control, service discovery, relationship management, service composition. |
|---|---|
| Architectural Layers | Foundation, Component, Application. |
| Key Studies and Findings | - Khan et al. [73]: Explored intelligent SIoT technologies to analyze human behaviors using EWM and TOPSIS methodologies, achieving effective behavioral assessments.<br>- Ullah et al. [74]: Proposed a hybrid approach with CFG and RNN for identifying duplicates, achieving 96.24% accuracy but lacked long-term sustainability insights. |
| Innovations and Frameworks | - Stelea et al. [76]: Developed models for virtual socially-aware objects but provided limited privacy implications analysis. |
| Controllability in SIoT | - Aghee et al. [77]: Defined controllability for temporal SIoT networks and proposed CSIoT to identify critical nodes, enabling distributed control. |
| Privacy and Security | - Bi et al. [78]: Introduced a framework using Bayesian game theory to balance QoS and privacy but did not address issues like incorrect access controls or software vulnerabilities. |
| Search SIoT (SSIoT) | - Fu et al. [79]: Introduced a two-tier SSIoT model based on social science theories, utilizing measures like average distance and network density but underperformed in real-world applications. |
| Clustered Design Architecture | - Abderrahim et al. [80]: Proposed TMCoI-SIoT for trust management using clustering but lacked clarification on recognizing and surviving attacks. |
| Advantages of SIoT Architectures | SIoT encourages collaboration and information sharing between people and devices, leveraging structured architectures like fog-based, cloud-based, and clustered designs. |

## 3. Current progress in IoT

The IoT seems to have an infinite number of possible applications. The industrial Internet will advance thanks to enhanced network strength, machine learning, and the capacity to create, run, integrate, and protect a range of applications. This is made possible by both enabling smart objects at once and collecting enormous amounts of useful data, which can be used to manage a variety of business operations. IoT apps run on IoT hardware and can be tailored for almost any industry, including hospitals, control systems, smart devices, vehicles, and smart wearables [81,82]. IoT applications increasingly employ AI and machine learning to provide the world with understanding and intelligence.

Smart information and communications technology (ICT) solutions can assist older populations with acute or chronic diseases by increasing their quality of life and allowing them to live independently. The massive data generated by IoT devices presents a significant challenge, as efficient processing and transmission become increasingly difficult. To address this, "Beyond Orion," a novel lossless compression method, was introduced by Mrewa et al. [83] to optimize data handling in IoT systems. The algorithm combined advanced techniques like Lempel Ziv-Welch and Huffman encoding with strategies such as pipelining, parallelism, and serialization to enhance efficiency and reduce resource usage. Experimental results demonstrated impressive data compression, achieving reductions of up to 99% across various datasets and a maximum compression factor of 7694.13. Comparative analysis further highlighted its effectiveness, with savings of 72% and a compression factor of 3.53.

Pateraki et al. [81] introduced biosensors as a dominant future of IoT, which are full analytical devices. They consist of a receptor that can detect the presence of a specific analyte. IoT has become a low-cost technology that offers modular and expandable solutions. Smart healthcare systems can regularly review the well-being and behaviours of individuals and the older population, along with the security and safety of the immediate environment, using wearables and ambient sensors. Their study also emphasizes how IoT is, in a way, linked to the Internet of Medical Things. Future IoT advances were illustrated using the SMART BEAR project, which aims to incorporate a perceptive and individualized technology platform for extending healthy and wholesome life by including a financially sustainable, stable, and discreetly innovative platform [81]. In future works, a more extensive application of biosensors in the real-world concerning data mining techniques and data analysis will be appreciated. Other concerns about the proposed biosensor include low processing capabilities, privacy protection systems, reliability, and immature security functions.

IoT business model innovation is a look into the future of IoT application and expansion. With a focus on crucial decision points and criteria, Tesch et al. [84] made one of the first attempts to identify new IoT-based business model breakthroughs and their dynamic models. The authors chose 13 instances from 8 businesses and selected case studies using the following criteria. The case firms are very prominent companies in the IoT ecosystem; the chosen project was a business strategy invention for the IoT context, and the interviewees were specialists in various IoT implementations of business roles and tasks. The results were obtained using the popular "Stage-Gate approach" for idea generation and the underlying basic notions of stages and gates [84]. The authors found that practitioners employ a repeating procedure, described by interviewees as a continual trial-and-error method, to build business models.

The main reason for this was increased complexity and ambiguity. The findings showed that, regardless of the kind of IoT business model innovation project, two gates appeared as important to go/kill or to prioritize decisions in all situations [84]. However, considering it was a qualitative case model, the study came with general limitations, including poorly detailed findings. Also, the authors failed to share the types of companies they focused on.

Jindal et al. [85] highlighted the future scopes of IoT and its challenges. The IoT has exploded in popularity and will continue to do so, yet any new technology brings uncertainty and business risk. The key requirements, such as hardware and software assets, are either in short supply or in development. It is also true that the security and confidentiality problems of IoT devices have not been adequately addressed during the last decade. The most critical difficulty in this regard is selective data storage on a cloud so that storage issues will not hamper future IoT device users. The key challenges in designing and developing IoT devices are privacy and security. Manufacturers and companies must consider the deployment of new security protocols as part of network layer security to ensure end-to-end delivery of sensitive data. Every facet of IoT, including technology, business, society, and law, works against its success. Therefore, acceptance of technology is also important and should be considered throughout its creation. People who do not like utilizing gadgets or smart devices or are uncomfortable interacting with technology would have a tough time working with the IoT's complicated capabilities [85]. Some limitations of the study are that it was confined to difficulties in business, technology, and society and failed to discuss the widely utilized IoT in other systems in detail.

Machine learning, particularly deep learning, assists individuals in almost every aspect of their lives, especially in mobile and IoT applications. Many defense methods, such as inefficient resource allocation and limited coordination capabilities, still encounter challenges. To address these issues, Feng et al. [86] introduced a novel adversarial security scenario and proposed a security game model that integrates defense resource allocation with patrol inspection. They designed a deep reinforcement learning algorithm called SDSA for this game model to determine the optimal security defense strategy. SDSA identifies the best patrolling strategy by exploring the policy in a multi-dimensional discrete action space and enables efficient cooperation among multiple defense agents by training a multi-agent Dueling Double Deep Q-Network (D3QN) with prioritized experience replay.

Zhang et al. [87] accentuated the current status of robust systems for machine learning-based applications, focusing on their challenges and certain research opportunities in deep neural networks (DNNs) efficiency and security. Challenges like permanent faults and timing mistakes arose as a result of process changes and aging. To decrease timing mistakes in MAC units of a TPU-like accelerator, the authors offered TE-Drop as a new approach. Inputs with carefully generated adversarial noise patterns were found susceptible to DNN-based classifiers [87]. Adversarial perturbations and backdooring attacks were two of the most hazardous threats. The paper also suggested countermeasures, such as pre-processing, trimming, and fine-tuning. All of the solutions, however, were confined to previously discovered attacks, and the study did not go into depth regarding the consequences of these assaults or how they influence the IoT infrastructure. Many machine learning-based systems are prone to various security and reliability issues.

The IoT has made significant advances in recent years, completely transforming how we use and experience technology and the physical world. Due to a growing number of connected devices, IoT has spread across many industries, from medicine to manufacturing to transportation to smart homes. Recent advances in networking technology, such as 5G and edge computing, have made it possible to analyze data in real-time at the very periphery of a network. In addition, IoT systems can now gain valuable insights from massive amounts of data owing to the incorporation of AI and machine learning. However, there are still challenges to overcome, especially when protecting the privacy and security of IoT devices and meeting the demand for interoperability. However, IoT is still developing rapidly, promising even more progress in the foreseeable future. Table 4 demonstrates the current progress in IoT.

Table 4. Current progress in IoT

| Aspect | Description |
| --- | --- |
| Applications | IoT has diverse applications in industries such as healthcare, smart homes, vehicles, wearables, and manufacturing, leveraging AI and machine learning to enhance functionality and efficiency. |
| Healthcare Impact | Smart technologies enable improved quality of life for older populations and individuals with chronic conditions by facilitating independent living through wearables, ambient sensors, and connected medical devices. |
| Data Handling | Advanced compression techniques and optimization strategies are necessary to handle massive IoT-generated data efficiently, reducing resource usage and enhancing data processing capabilities. |
| Biosensors | Biosensors play a crucial role in IoT monitoring health and environmental conditions, with ongoing challenges related to processing capabilities, privacy, reliability, and security needing further exploration and improvement. |
| Business Models | IoT-driven business models focus on innovation and iterative development processes, highlighting the importance of adaptability and strategic decision-making in dynamic markets despite challenges like limited transparency in model design. |
| Challenges | Key IoT challenges include addressing security and privacy concerns, managing selective data storage, overcoming hardware/software limitations, and ensuring user acceptance of IoT's complex functionalities. |
| Security Innovations | Effective IoT security requires advanced models integrating resource allocation and strategic defense mechanisms, enabling efficient coordination among security systems to counter threats. |
| Machine Learning | IoT systems face challenges in deploying robust machine learning models, including vulnerability to adversarial attacks and aging-related faults, with proposed solutions focusing on targeted countermeasures and reliability improvements. |
| Technological Advances | Innovations like edge computing and next-generation networking allow real-time data processing at the network edge, while AI integration provides actionable insights from massive IoT data streams. |

| Future Prospects | IoT is evolving rapidly, with promising developments in interoperability, enhanced security, and expanded applications across industries, though challenges in scalability and privacy remain to be addressed. |
|---|---|

## 4. Forensics and security issues in IoT

As the number of connected devices in use grows, forensics and security concerns in the IoT are more important than ever. Since IoT is a massive network of linked devices that can interact with one another [88], share information, and carry out a wide range of useful tasks, this interdependence poses serious issues for forensics and safety. One important issue is that IoT devices are typically unprotected and open to cyberattacks because they lack proper security protocols. As a result, unauthorized users could gain entry, compromise data, or even take control of vital infrastructure.

As with any new technology, there are advantages and disadvantages to using blockchain technology for IoT forensics and security. Due to the large volume of transactions caused by IoT devices, scaling blockchain networks can be challenging. The restricted computing power and storage space of IoT devices complicate blockchain implementation [89]. In addition to putting pressure on blockchain networks, the large data volume created by IoT devices also necessitates careful selection and transfer of user data. The inherent delays in blockchain consensus systems may conflict with the need for low latency and near-instantaneous responses. The immutability of blockchain transactions threatens privacy and confidentiality [90]. Integrating blockchain and IoT presents regulatory and governance concerns that necessitate updating legal frameworks. Consensus algorithms in networks can be very resource-intensive. Exploring blockchain architectures designed to the specific forensics and security requirements of the IoT is essential for finding solutions to these issues.

### 4.1 Forensic issues with IoT

IoT forensic issues include the wide variety of concerns forensic investigators encounter while examining IoT gadgets and systems [91,92]. The wide range of IoT devices is a major issue, including anything from smart home appliances and wearables to industrial sensors and autonomous vehicles [93]. Each device might have its unique combination of hardware, including but not limited to its operating system, data formats, applications, and network protocols.

Therefore, forensic investigators must deeply understand the devices to assess and collect evidence from them properly. The lack of uniformity in data formats, encryption, and device configurations further complicates forensic investigations [94]. Additionally, the sheer scale and real-time nature of IoT systems make it challenging to identify and preserve crucial evidence promptly. These complexities necessitate specialized tools and techniques to address the unique forensic needs of IoT environments.

There are many situations (Fig. 3) in which IoT forensics is required. Still, the most common ones involve investigating and resolving security events, breaches, or other malicious acts involving IoT devices and systems [95,96]. IoT forensics is essential to investigate security issues, guarantee accountability, ensure privacy, avoid future attacks, and assist in legal procedures. It is a crucial field that ensures the security and reliability of IoT infrastructure across many sectors. Investigating, analyzing, and preserving digital evidence from electronic devices, including computers, cellphones, and storage media, is the subject of digital forensics, a subfield of forensic science [97]. Using scientific methods and tools is essential for collecting, evaluating, and interpreting digital evidence supporting investigations and legal procedures. There are a number of key differences and issues with IoT forensics related to the forensics of more conventional devices [95] in terms of legal and regulatory frameworks, forensic tool availability, privacy and consent, data volume and velocity, time synchronization and timestamping, security and encryption, and fragmented evidence.

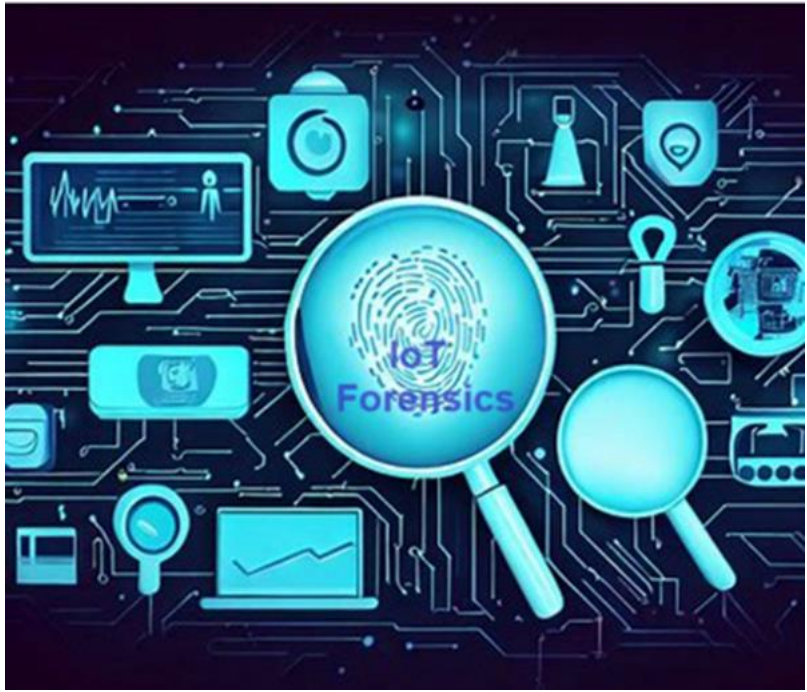

Fig. 3. A schematic for IoT forensics

IoT involves interconnected nodes that communicate with each other independently while providing a diverse range of facilities. Yet, this open system's extensive deployment and enormous computing power make IoT devices an ideal target for various cyberattacks. Especially if the nature of the information that is collected and processed over an IoT network is private, it becomes even more susceptible to malicious attacks [16]. This is why ensuring the security and ways to identify nodes within a network that have been compromised is a huge priority for IoT networks. The reliability and convenience that IoT brings to the table also open a dimension of intruder entrance points. This leads to the area of digital forensics, which not only identifies compromised network objects but also retains evidence of attacks [34]. Forensics also allows developers to build better protection systems in IoT networks for future attacks.

*4.1.1 Evidence identification, collection, and preservation*

IoT has escorted a technological evolution that also introduces a new kind of threat for cyber-security concerns, given the complex and heterogeneous nature of IoT objects as well as the unique susceptibilities of IoT networks. It is necessary to detect devices that have been compromised and to access and conserve evidence of malicious activities that can emerge as high-priority events.

There have been many attempts to study processes of evidence identification, collection, and preservation within IoT networks. For instance, a study targeting this problem used a blockchain-based solution designed for domains such as smart homes, explicitly collecting and preserving valuable evidence in digital forensics [22]. The study used a database that stores private forensic evidentiary data where the accumulated evidence would be kept, coupled with a permitted blockchain that would provide facilities, such as integrity and authenticity, to ensure courts can utilize this evidence. This method saves metadata of evidence that would be extremely useful for providing appropriate services. It allows it to communicate with various processing objects, such as ISPs, law enforcement agencies, and other entities. This approach allows for handling very different challenges associated with forensic evidence from IoT networks. However, many drawbacks of blockchain technology, such as scalability, open nature, and complexity, could cause inefficiency and overkill of the proposed approach.

A generic methodology was developed by Alqahtani & Syed [98], integrating each forensic transaction into an immutable blockchain entry, which established transparency and authenticity from data preservation to final reporting. The framework was designed to manage a wide range of forensic applications across various domains, including technology-focused areas such as the Internet of Things (IoT) and cloud computing, as well as sector-specific fields like healthcare. Central to the approach were smart contracts that seamlessly connected forensic applications to the blockchain via specialized APIs. A privacy-preserving blockchain-based scheme for supply chain traceability was proposed by Li et al. [99]. The approach leveraged zero-knowledge proof technology, allowing participants to enter the system using a one-time pseudonym, thereby protecting their identity information. A secure and reliable method for sharing product information was proposed, utilizing attribute encryption technology to encrypt product information stored in the blockchain.

Hossain et al. [100] developed a framework that uses a public digital ledger to detect the evidence of criminal activities in IoT networks. Due to increased deployment, IoT devices have become easy targets for certain attacks. This approach identifies and collects interactions in IoT networks and objects as evidentiary data, then stores them in a secure system similar to a Bitcoin network. This eliminates a single entity over-storing evidence, helps avoid single-point-of-failure on the storage media, and ensures that evidence is highly available. The framework also assures integrity, confidentiality, and anonymity of evidence. However, as discussed before, it also has drawbacks similar to those of blockchain technology.

A study by Bouchaud et al. [101] discusses three key steps in evidence gathering and preservation in IoT: detection, localization, and recognition. The detection mechanism is often achieved passively through a frequency mapping that recognizes the signature at the scene as a digital footprint. The scan is done at several layers and, thus, detects object presence at the crime scene. One challenge of this approach is that only active devices are detected, which is time-consuming and sometimes incomplete. Evidence localization is achieved by studying RSS (received signal strength) and performing many measurements to locate devices, which provided a direction in the investigation. Finally, recognition is conducted at two levels: i) digital recognition is done in numerous ways, such as collecting MAC addresses of communicating devices to generate a trace, and ii) physical recognition is done by utilizing the electromagnetic signature that depends on the time for response and signal confirmation to characterize the device's physical location uniquely. It can also provide information about the device's manufacturer and country of origin.

When conducting forensic investigations involving the IoT, it is crucial to correctly identify, gather, and store all evidence. The first step is to secure the crime scene and isolate the associated IoT devices to preserve any potential evidence. Information on the device's capabilities and vulnerabilities can be better understood with complete documentation of the device's development, model, and firmware versions. Information on network activity and potential security breaches can be gathered by collecting network and device data, such as logs and configurations.

The correct extraction, analysis, and interpretation of evidence require specialized forensic tools and the cooperation of experts in other fields, such as digital forensics and IoT security experts. Accurate and valid investigation outcomes in IoT forensic cases are ensured by adhering to established forensic processes and requesting professional guidance. If possible, preserving cloud data allows users to reconstruct activities and events conducted in the cloud. Data synchronization between gadgets and systems should be carefully considered. Evidence obtained must be kept secure and legally admissible, which can be achieved by maintaining a chain of custody records.

*4.1.2 Evidence analysis and correlation*

IoT forensic investigations rely heavily on analysis and correlation of evidence to reconstruct what happened and provide conclusions about its significance [102]. Analyzing network traffic captures, device logs, and forensic investigation of firmware or software are some methods that can be used to analyze the obtained data [103]. This analysis focuses on patterns, deviations, and possible evidence of access or malicious behavior.

Digital evidence in IoT is incredibly important as it makes the case of cybercrime admissible in a court of law. The complex system of an IoT network includes collecting, analyzing, and correlating evidence, which can be achieved by many relevant forensic methods and tools fruitfully and efficiently. One study utilized the Last of Scene (LoS) algorithm for evidence collection and analysis, investigating the last item found in communication first [26]. This facilitated and lowered the odds for the investigators in digital forensics. LoS can be easily explained by limiting the investigation to a 0 zone before any future workload and overhead occur. This procedure was more efficient because of its ability to identify several objects and use otherwise hard-to-reach entities.

The versatility of IoT devices makes them much more susceptible to continuous cyber-attacks. Additionally, it is difficult for forensic investigators to keep records of many attacks on these networks due to the lower processing power and memory of IoT [104]. Another study proposed an intelligent forensic analysis mechanism that automatically identifies an attack on an IoT network utilizing a machine-to-machine (M2M) framework. Here, the evidence collection and analysis are performed using a third-party logging server to determine the effects and nature of the attacks. However, since the process adopts a machine learning framework, it has many drawbacks of ML, such as high training datasets and high error vulnerability.

The increasing complexity of multi-level interconnections in IoT ecosystems has introduced new challenges in identifying connected devices. Kim et al. [105] examined IoT services capable of connecting to various devices to reveal their interconnections, analyzing forensic artifacts across six interconnected scenarios. The study found that not all services retain traces of interconnection. Based on the findings from scenario-based experiments and existing IoT forensic frameworks, an improved IoT forensic model was proposed to identify the interconnectivity between devices. The framework introduced additional phases for identifying interconnected devices, designed to investigate IoT-related cases by identifying all connected devices, ensuring relevant data collection, narrowing down areas of investigation, and integrating the data. The proposed model facilitated the discovery of interconnected devices.

Another study introduced a fog-based IoT forensic framework that tries to acknowledge and work on the challenges in evidence analysis [24], in which resources are distributed to the edge of a network for computing and storage. This approach showed improvement in scalability, lower latency of the network, quick responses, and better security and privacy than other methods in the literature. Some drawbacks of fog computing include complexity, higher risk of attacks due to increased connections, data leaving premises, and more encryption, which makes it harder to share data across networks. A study by Kebande et al. [106] introduced a cloud-centric protocol to isolate big data for forensic evidence from IoT infrastructures for accurate analysis and investigation. However, integrating the cloud with IoT has created a massive array of data that is open to malicious risks from cybercriminals. The authors proposed a CFIBDIoT framework where evidence was framed to be collected over IoT clouds. Results show that the framework can support a cloud-based IoT tool with good accuracy.

In conclusion, matching timestamps, comparing network activity with device logs, or lining up activities made on IoT devices with cloud platform data are examples of how to correlate the evidence. This correlation method helps determine dates, establish cause-and-effect relationships, and assign responsibility for events to particular tools or people. Accurately interpreting the evidence and discovering valuable insights calls on knowledge of digital forensics, network analysis, and IoT security. Working collaboratively with legal professionals guarantees that the evidence analysis and correlation meet all legal criteria and may be presented competently in legal proceedings. An in-depth understanding of the IoT forensic issues at hand can be attained by investigators through careful examination and correlation of the available information.

### *4.1.3 Attack or deficit attribution*

A common outcome of any investigation in forensics is to find the criminals involved in the attack. However, given the rise in the autonomous vehicle industry, it will soon become a problem for forensic investigators to identify perpetrators of cybercrimes. This is due to the lack of documentation and sound tools of forensics to collect, preserve, and analyze cyber-physical systems. Without a proper system, this becomes a challenge [16]. Sana et al. [107] focused on anomaly detection, which identifies deviations from normal system behavior that may indicate attacks. The study explored methods to enhance detection performance, investigating the design and evaluation of a novel IDS. Supervised ML techniques, including tree-based Support Vector Machines (SVM), ensemble methods, and neural networks (NN), were leveraged alongside advanced deep learning approaches such as long short-term memory (LSTM) and vision transformers (ViT). Singh [108] used a cross-layer design as a modern architectural method to establish communication between layers that can adapt quickly. Results indicated that the technique helped provide quality service, lowered energy consumption, and encouraged efficient resource utilization.

Limiting attacks or detecting gaps in IoT forensic issues can be difficult and require a well-thought-out study. Attack attribution determines who or what was responsible for an attack [109]. This could be an external hacker, an insider, or a misconfiguration. Expertise in digital forensics, network analysis, and cybersecurity is needed to determine where an attack came from, how it spread digitally, and what vulnerabilities were exploited. When deficiencies must be determined, attention moves to analyzing the security issues in the IoT device or system that enabled the breach [110]. Methods of authentication, firmware releases, and update management policies should also be examined. Understanding the attack methods, motivations, and possible countermeasures requires cooperation with IoT security professionals and using threat intelligence.

Experts in forensics, IoT development, networking, and security must work collaboratively to solve the identified forensic issues. This process includes implementing privacy-aware protocols, improved data integrity procedures, and IoT system-wide privacy awareness. By addressing these issues, the forensics community can better protect the privacy, security, and reliability of IoT devices and networks, benefiting the entire IoT ecosystem.

## 4.2 Security issues with IoT

While the IoT has greatly improved our experience with technology, it has also raised a number of security issues. The connected devices' inherent weaknesses are a significant source of concern when it comes to IoT security [111,112]. Due to the constraints of computing power with memory, many IoT devices do not have adequate security features [113]. Because of this, they may be open to attacks and misuse. Furthermore, malicious people can more easily obtain control of devices or access sensitive data due to poor authentication and authorization methods in IoT devices. Another significant issue is insecure communication, as many IoT devices use wireless transmission of data, leaving it open to eavesdropping, malicious activity, and tampering.

The IoT enables uncountable devices, users, and services to communicate and exchange data. It is applied in smart cities, such as highways and hospitals, and smart homes, such as operating doors, air conditioner units, and fire alarms, among other things [114]. These IoT gadgets are linked to the network via the Internet and exchange data through that network [115]. Due to the rising use of IoT devices in multiple sectors, IoT networks are prone to various security vulnerabilities [116]. For example, data obtained from IoT-embedded sensors are delivered to cloud servers or other analytical engines via middleware, like gateways and routers. These IoT-processed data and information have garnered a great deal of positive and negative attention. Several unexpected security challenges for current networking technologies have arisen because of the inherently open IoT structure and widespread adoption of the paradigm [117]. The security problems in IoT vary, observed in individual authentication, authorization, access control, and data protection, as well as in the architecture, as shown in Fig. 4.

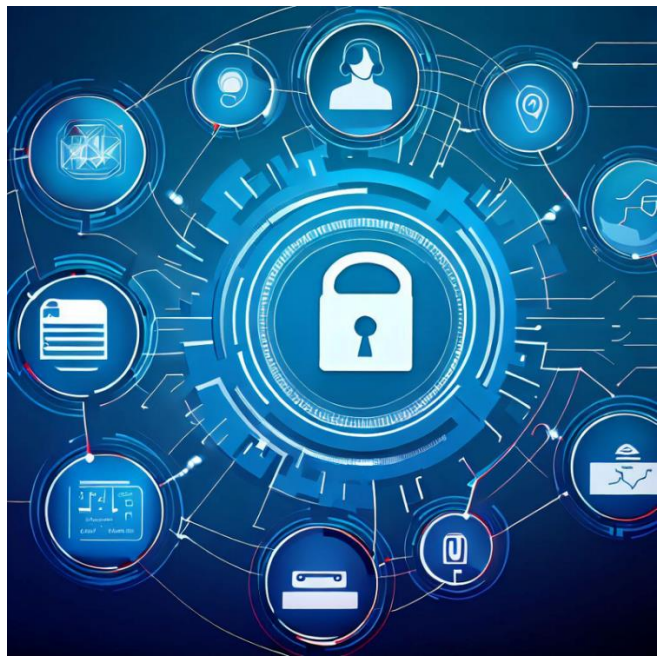

Fig. 4. Schematic for security issues in IoT

*4.2.1 Authentication*

When it comes to protecting data on the IoT, authentication is essential. Authenticating individuals and verifying the authenticity of their devices has taken on more importance as the number of connected devices continues to grow [118,119]. To ensure that only authorized people, machines, and organizations have access to the infrastructure and services of the IoT, authentication procedures are used. Credentials like biometric data, digital certificates, passwords, and multifactor authentication mechanisms are all part of this process.

Every object in the IoT should be able to recognize and authenticate every other item within the network, or at least in a specific area where the system would interact [120,121]. Verifying devices and users before accessing a network or a communications system is known as authentication. Authenticating each gadget is a tedious task that should be resolved by the specified architecture [122]. Authentication occurs in secured and unsecured networks, making them susceptible to malicious assaults, such as distributed password dictionary attacks, denial-of-service (DDOS), and man-in-the-middle attacks [123]. Authentication is challenging due to the heterogeneity of linked devices and multiple security schemes. Traditional techniques that focus on usernames and passwords or involve a single authentication pattern are perhaps insufficient for coping with the heterogeneity expected by IoT as they are readily exploited [124,125]. Any proposed authentication strategy must have a lightweight feature for many IoT devices with low-power CPUs, minimal computation, and storage. Additionally, the system may implement multi-factor authentication that entails multiple authentication factors simultaneously. It must efficiently handle the complementary load of multi-factor authentication and may integrate encryption technology, such as RSA, Elliptic Curve, AES, or hash functions, to ensure robust security [126,127].

To address authentication concerns, researchers developed various methods tailored to the IoT context [125]. For instance, a novel lightweight authentication mechanism, named RAM-MEN, was introduced by Tanveer & Aldossari [128] to secure IoT-enabled Mobile Edge Computing (MEC) environments in the 6G era using cryptography and physically unclonable functions. The mechanism protected against insider threats and fake MEC access points while ensuring efficiency and scalability. RAM-MEN also established a secure communication channel (session key) between IoT devices and the MEC server, enabling the secure offloading of computationally intensive tasks. The security of the session was rigorously evaluated through formal methods, including Scyther and the random-or-real model, as well as informal approaches. Comparative performance evaluations demonstrated that RAM-MEN reduced communication costs by 21.54% to 45.53% and computational costs by 17.09% to 83.72% while providing enhanced security features. Li et al. [124] presented a blockchain-based authentication mechanism rather than relying on a third party to mitigate the pitfalls and assaults that might develop when counting on a centralized authority. This method is lightweight, affordable, and tamper-resistant as it depends on public-key cryptography. Besides, it eliminates single-point failure attacks and ensures the system continues functioning effectively even if some nodes are targeted by a DDoS attack. Moreover, the multi-chain structure provided additional security protection between different trust domains. However, its performance depends on the blockchain platform. Subsequently, Shah and Venkatesan [129] developed a mutual authentication technique that uses a shared secret key and multi-password between the IoT device and server. This approach shields against side-channel attacks, password cracking, DOS assaults, and man-in-the-middle attacks. The set of passwords must be updated after every session, making it resistant to several attacks even with limited power. The method was tested on an Arduino device to verify its feasible implementation with constraint memory and computational capability on IoT devices. However, the key size must be extended to enhance security, which would necessitate high processing and more energy.

Authentication of IoT systems associated with multi-WSNs is a novel decentralized strategy to fix the problems that cause centralized authentication [130–132]. A public blockchain-based multi-wireless Sensor Network (WSN) verification method was proposed by Anitha & Bapu [133] using a Light-Weight Authentication Algorithm (LWAA) for IoT to enhance secure authentication and performance in a multi-WSN model. This approach categorized nodes as access points, group head nodes, and regular nodes based on their power variations, forming a hierarchical model. The standard authentication of nodes across various communication scenarios was managed through blockchain. The cryptographic technique improved the network lifespan and significantly reduced computation time. Khalid et al. [134] introduced a decentralized authentication concept for multiWSNs' IoT-based communication infrastructure using fog computing as an open authentication mechanism to validate the credibility of cooperating sensor devices. This scheme outperformed the state-of-the-art blockchain-based authentication system. Nevertheless, further study might be conducted to create a lightweight consensus protocol to prevent massive energy consumption using the Proof-of-Work (PoW) algorithm to evaluate every block. Radiofrequency identification (RFID) can be significantly favorable to identifying entities, as it uses electromagnetic wave propagation and induction to differentiate across objects. Cryptographic, password-based, biometric, token-based, zero-knowledge proof, digital signature, and multi-factor authentication are some security mechanisms experts recommend addressing the same issue. Alternatively, researchers are particularly interested in multi-factor authentication because of its adaptability in rendering the authentication process safer and more effective [135–139].

In summary, unauthorized access, data breaches, and malicious actions can be prevented with the help of strong authentication systems. As the number of IoT systems and vulnerabilities to these systems rise, authentication strategies thus need to be optimized, which necessitates consistent modifications to current procedures. Authentication also aids in keeping private and confidential data transmitted inside IoT ecosystems secure. In order to establish reliability, prevent illegal access, and preserve the integrity of IoT security, strong authentication techniques, in addition to secure key management and communication protocols, are required. For IoT security measures to be reliable and successful, authentication technologies and standards must keep improving to implement them correctly and educate users.

*4.2.2 Authorization and access control*

The IoT ecosystem relies heavily on authorization and access control to solve security problems [119,140]. With so many connected devices, it's critical to restrict access to important data and features to authorized users only. IoT systems use efficient authentication and access control technologies to manage user access and prevent unauthorized behavior [141]. This entails developing and implementing policies that specify which users and devices have access to which data and services and under what conditions.

With the rise of IoT, prevalent use is responsible for creating a mass number of log information that is transferred, shared, and processed regularly. These logs might contain sensitive and necessary user information and might hold evidentiary details in forensic investigations. However, the legal aspects of this evidence and internal safety issues in current logs are not yet adequately understood. Hsu et al. [142] proposed a log storage management system that autonomously adheres to a blockchain mechanism and access control for IoT services. This model allows sensors to log encryption when sending to the gateway and server, so it would not be publicly readable if assessed between communications. In addition, a concept of "signature chain" was also introduced that provides efficiency in management with valuable security assurance for logs, including verification of identity, data integrity, non-repudiation, resistance to data temperament, and legality.

Due to the protocol's use of blockchain technology, weaknesses associated with blockchains, such as scalability, open nature, inefficiency at certain times, and complexity, can pose problems. IoT is also becoming highly prevalent in healthcare sectors, particularly smart sensors and wearable devices that make medicine smart, fast, and accessible. However, medical data makes IoT far more susceptible to attacks and raises a big concern for security and users' access control. Considering the increasing size and consistent presence of this system, a more manageable protocol that is flexible and scalable is required. A novel Ciphertext Policy Attribute-Based Signcryption (CP-ABSC-MIoT) scheme was proposed by Patil [143] to provide privacy preservation and access control for Electronic Health Records (EHR) in Mobile Internet of Things (MIoT) environments. The scheme aimed to achieve fine-grained access control, confidentiality, authenticity, and privacy preservation of EHRs by combining the advantages of attribute-based encryption and digital signatures. The system enforced access policies tailored to the unique characteristics of authorized EHR users. The proposed scheme, built on bilinear pairing, supported several security features, including data confidentiality, authentication, and efficient access control. Pal et al. [144] proposed an architecture for access control in constrained healthcare environments that use IoT resources. The policy-based approach provides neatly analyzed access for services to authorized users while protecting important data from intrusion. In addition, a hybrid route is used for authorization design, which employs roles, attributes, and capabilities. It applied parameters for the role membership assignment and evaluating authorization. The membership permits the assignment of capabilities based on additional user attributes, which is then used to grant access permission to specific IoT-related services. This lowered the policy number needed for user access control specification settings. This particular framework is XACML-oriented and implements a prototype known as proof-of-concept and analysis. The assessment results proved that this approach needs less overhead in comparison to existing schemes for IoT's access control issues.

There are some drawbacks to using XACML, such as not explicitly requiring specification for the intent needed in privacy policies, complexity, and not having standardized interactions involving PAP and PIP. Sadineni et al. [145] presented a holistic forensic model based on ISO/IEC 27043 international standard containing three phases corresponding to proactive, incident, and reactive that covered the full of IoT forensics: preparedness, initialization, and inquiry. The model provides a configurable and customizable environment that would support a diverse range of IoT applications and would be better to build a comprehensive framework.

Overall, verifying the identities of people and devices is made easier by strong authentication protocols like digital certificates and multifactor authentication. Fine-grained management of permissions can be exercised on the basis of user roles, device attributes, or other contextual elements with the help of access control techniques like role-based access control and attribute-based access control. Authorization and access control systems are more effective in preventing data breaches, malicious actions, and unauthorized access when properly implemented and configured. The integrity and security of IoT networks, as well as the protection of sensitive information, depend on the regular monitoring, auditing, and updating of access control policies.

### *4.2.3 Privacy*

IoT privacy safeguards one's personally identifiable information from disclosure in IoT-protected space. An individual is given a unique identifier and the ability to communicate on the Internet in an IoT-protected environment. One of the best features of IoT is that it offers consumers a significant amount of operation and control over routine tasks by filling the surroundings with smart devices. Security and privacy have become the main concerns because of the overwhelming usage of IoT devices [146]. Limitations of devices and complex heterogeneity significantly impact IoT privacy, as IoT devices are often limited with constrained resources because of weight and size and network access issues.

According to Yu et al. [147], IoT devices can be exploited to disclose confidential information by providing vulnerable access points to key facilities, such as healthcare and defense institutes. Two key findings regarding IoT systems are: i) host-based techniques are more vulnerable than network-based approaches because of the underlying restrictions and plausible unfixed security flaws, and ii) statistic perimeters are implemented deep within the IoT network due to their constant physical and cyber-shifting, these are incapable of safeguarding IoT technologies. As a result, resource constraints make each IoT layer's security difficult [148].

A significant risk is posed by the flow of data produced by IoT devices, which may breach people's privacy and compromise their security [149]. One of the most serious misuses of the massive amounts of data exchanged is when that data is connected to a particular individual. As a result, user obfuscation is another factor to consider when maintaining confidentiality in networked environments. Miorandi et al. [150] explain using a suitable query language to allow apps to retrieve the required data and maintain secrecy in an IoT situation. Usman et al. [151] proposed a security infrastructure that considers user anonymity, security problems, and confidentiality. The author prioritized IoT security and addressed the challenges of IoT manageability, availability, and privacy.

In summary, protecting users' personal information is a top priority in IoT security. Data privacy and security are major issues brought to light by the massive amounts of data networked IoT devices collect. Adopting strong privacy protections to protect user privacy is critical as IoT devices continuously capture personal and sensitive information. This involves embracing privacy-by-design principles, which call for incorporating privacy protections into every step in developing an IoT system. Data confidentiality can be protected through the use of anonymization and encryption methods when being sent or stored. Users should be presented with clear and transparent privacy rules that explain the collection, usage, and sharing of their data. Individuals are given more agency over their data when given choice and control over it through mechanisms including gaining informed consent from users and having access to their data. Potential privacy risks in IoT systems should be identified and remedied through routine security assessments and audits.

### *4.2.4 Secure architecture*

A secure architecture is required to ensure strong security within the IoT ecosystem [152,153]. To secure IoT networks, it is necessary to develop and deploy a multi-pronged, defense-in-depth strategy adapted to the specific threats and weaknesses of these networks. The term "security architecture" refers to a comprehensive approach to securing a given situation or environment that takes into account both the requirements and potential risks. A novel security framework was introduced by Lilhore et al. [154], leveraging deep learning to address risks in IoT channels and 5G networks. The framework utilized lightweight CNNs (MobileNetV3-SVM) and transfer learning to detect intrusions in real time by analyzing network activity patterns and identifying malicious behavior efficiently. It employed a multi-layered structure to hierarchically learn from raw network data, reducing computational overhead and making it suitable for IoT and 5G edge devices. The hybrid model combined MobileNetV3 with SVM for auto-classification, enhancing intrusion detection accuracy and adaptability to evolving threats. Koroniotis et al. [155] proposed a network forensics framework implementing Deep Learning for IoT that uses Particle Deep Framework (PDF), which describes the phases of a digital investigation for proper identification and tracing of attacking behaviors within IoT networks. The framework includes the following three novel functionalities: i) extraction of network flow data and verification of their integrity to deal with encrypted networks; ii) utilization of particle swarm optimization (PSO) algorithm to automatically adapt to parameters of deep learning, and iii) development of a deep neural network based on the PSO algorithm to find and trace events laced with abnormality from smart home IoT networks. The framework was evaluated using Bot-IoT and UNSW_NB15 datasets and compared with other deep learning algorithms. Results implied a high-performing system for the discovery and tracking of events of cyber-attacks in comparison to other techniques. A few drawbacks of implementing a deep learning network include requiring large amounts of data for better performance, expensive training modules for complex data models, and complexity overall.

How blockchain technology was used to create a tamper-proof and transparent system for secure data sharing in autonomous vehicular (AV) networks was analyzed by Hemani et al. [156]. Smart contracts, implemented in Solidity and deployed on a blockchain using Ganache and Truffle, established predefined rules for data sharing. The proposed approach enhanced the security, safety, and reliability of AV networks, with MATLAB used for system analysis. A blockchain-based shared audit framework (BSAF) was proposed by Shakeel et al. [157] for analyzing digital forensics data from IoT. The framework was created to detect and trace the causative agent of data scavenging attacks in virtualized resources (VRs). Blockchain technology was used for accessing logs and managing control. The access log was analyzed for consistency in detecting events using logistic regression (LR) machine learning and cross-validation. Any adversary event discovered was filtered after cross-validation to hold back the precision level of data analysis for a variety of user densities and VRs. Results showed that the framework is consistent and reduces both analysis time and adversary event rate. Drawbacks of machine learning, such as high levels of error susceptibility, time-consuming, and resource consumption, as well as blockchain technology, like scalability, open nature, and complexity, may pose problems within this framework.

In general, there are a few foundational components of an IoT-safe architecture. There have to be robust procedures in place for authenticating devices and users and controlling access to resources. To protect information while it is stored and in transit, encryption mechanisms should be used. To avoid data tampering or interception, secure communication routes, such as encrypted tunnels and secure protocols, should be set up. By isolating and organizing networks, attacks can be contained and their damage minimized. Addressing security issues and protecting against known exploits requires regular software updates and patch management techniques. It is easier to notice and respond to potential threats or unusual behaviour with the help of intrusion detection and prevention systems and effective monitoring and logging procedures. In order to successfully resolve security issues and limit damage, a complete incident response strategy should be established. Therefore, IoT systems can reduce risks, prevent attacks, and retain ecosystem trust by employing a secure architecture. Table 5 summarizes some recent studies that investigated forensic and security issues of IoT.

Table 5. Forensic and security issues explored/discussed in IoT studies

| Issues | Issue types | Objective | Outcome | Applications | Remarks | Ref. |
|---|---|---|---|---|---|---|
| Forensic issues in IoT | Evidence identification, collection, and preservation | Handling very different challenges coming from forensic evidence from IoT networks. | By using evidentiary data where the collected evidence would be kept coupled with a permitted blockchain that would provide facilities such as integrity and authenticity, making sure courts can utilize this evidence. | Interact with different objects in the processing such as ISPs, law agencies, and other entities. | Scalability, open nature, and complexity could cause inefficiency and overkill. | [84] |
| | | Integrate forensic transactions into immutable blockchain entries for transparency and authenticity in forensic processes. | Designed a blockchain framework with smart contracts and APIs to manage forensic applications across diverse domains. | Applicable in IoT, cloud computing, and healthcare forensics to enhance data preservation and reporting authenticity. | Provides a versatile approach to forensic management but may require testing for scalability and domain-specific needs. | [98] |

| | | Develop a privacy-preserving blockchain-based scheme for supply chain traceability and secure information sharing. | Proposed a system using zero-knowledge proofs and attribute encryption for secure identity protection and encrypted data sharing. | Applicable in secure supply chain management to protect participant identities and ensure reliable product traceability. | Demonstrates a secure traceability model but may face challenges in large-scale implementation and computational efficiency. | [99] |
|---|---|---|---|---|---|---|
| | | Detect the evidence of criminal activities in IoT networks. | Allowed the elimination of a single entity over storing evidence, helped avoid single-point-of-failure on the storage media, and ensured evidence was highly available. | Store evidentiary data in a secure system similar to a Bitcoin network. | Similar drawbacks to blockchain technology include scalability, open nature, and complexity. | [85] |
| | Evidence analysis and correlation | Followed a LoS (Last of Scene) algorithm for evidence collection and analysis. | This algorithm investigated the last item found in communication first. | Investigation in digital forensics. | This procedure was more efficient because of its ability to identify a number of objects and use otherwise hard-to-reach entities. | [37] |
| | | Address challenges in identifying interconnected devices within multi-level IoT ecosystems by | Proposed an improved IoT forensic model incorporating additional phases for identifying | Applicable in IoT forensics to discover connected devices, collect | The framework effectively identifies device interconnectivity but may require further refinement to | [105] |

| | | analyzing forensic artifacts. | interconnectivity and investigating IoT-related cases. | relevant data, and streamline the investigation process. | handle large-scale IoT ecosystems. | |
|---|---|---|---|---|---|---|
| | | Automatically finds out the attack done on an IoT network utilizing a machine-to-machine (M2M) framework. | The evidence collection and analysis were done using a third-party logging server to determine the effects and nature of the attacks. | Investigation in digital forensics. | High training data sets and high error vulnerability. | [88] |
| | | Fog-based IoT forensic framework that tries to acknowledge and work on the challenges in evidence analysis. | This approach showed improvement in scalability, lower latency of the network, and quick responses. | Evidence analysis | More encryption makes it harder to share data across networks. | [36] |
| | | Enhance anomaly detection by developing a novel intrusion detection system (IDS) using supervised machine learning and deep learning techniques. | Demonstrated improved detection performance through the integration of tree-based SVM, ensemble methods, neural networks, LSTM, and ViT. | Applicable in IoT security for detecting system anomalies, preventing attacks, and securing networked environments. | The study presents a robust approach to anomaly detection but may face challenges in scalability and adaptability to diverse IoT scenarios. | [107] |
| | Attack or deficit attribution | Identify the perpetrators of cyber-crimes. | Used a cross-layer design, a modern architectural method useful to establish communication between layers and adapt very easily. | Identify the perpetrators of cyber-crimes in the autonomous vehicle industry. | Lowered the consumption of energy, and encouraged efficient resource utilization. | [90] |
| Security issues in IoT | Authentication | Blockchain-based authentication mechanism. | It eliminated single-point failure attacks and assured that the system functions correctly even though a DDoS attack targeted certain nodes. | Mitigate the pitfalls and assaults that might develop when counting on a centralized authority. | The multi-chain structure provided additional security protection between different trust domains. | [102] |
| | | Develop a lightweight authentication mechanism, RAM-MEN, to secure IoT-enabled Mobile Edge Computing (MEC) environments in the 6G era. | Ensured protection against insider threats and fake MEC access points, reduced communication costs by 21.54%–45.53%, and computational costs by 17.09%–83.72%. | Secure IoT communication in MEC environments for the 6G era, enabling safe task offloading and efficient resource utilization. | Demonstrated robust security through rigorous evaluation, but further testing in diverse real-world scenarios may be beneficial. | [128] |
| | | Introduce a public blockchain-based multi-WSN verification method | Enhanced secure authentication and performance in multi-WSN models, reduced computation | Secure IoT authentication in multi-WSN setups | The hierarchical model improves scalability and efficiency, but privacy | [133] |

| | | with a Light-Weight Authentication Algorithm (LWAA). | time, and extended network lifespan. | for applications in monitoring, environmental sensing, and smart grids. | concerns and resource constraints could be explored further. | |
| --- | --- | --- | --- | --- | --- | --- |
| | | A mutual authentication technique that used a shared secret key and a multi-password between the IoT device and the server. | It shielded against side-channel attacks, password cracking, DOS assaults, and man-in-the-middle attacks. | Authentication | The key size must be extended to enhance security, which would necessitate high processing and more energy. | [108] |
| | | A decentralized authentication concept for multiWSNs' IoT-based communication infrastructure using fog computing. | This scheme outperformed the state-of-the-art blockchain-based authentication system. | Validate the credibility of cooperating sensor devices. | Further study might innovate a lightweight consensus protocol to prevent the massive energy consumption by the Proof-of-Work (PoW) algorithm. | [113] |
| | Authorization and access control | An autonomous log storage management system adhering to a blockchain mechanism | This model allowed the sensor to log encryption when sending to the gateway and server. | Log storage management system. | A concept of a "signature chain" is also introduced that provides efficiency in management with valuable | [122] |
| | | and access control for IoT services. | | | security assurance for logs, including identity verification. | |
| | | Develop the CP-ABSC-MIoT scheme for privacy preservation and fine-grained access control of Electronic Health Records (EHR) in MIoT environments. | Achieved confidentiality, authenticity, and privacy preservation for EHRs through attribute-based encryption and digital signatures. | Secure management of EHRs in healthcare systems, ensuring controlled access and data privacy in Mobile IoT settings. | Strong security features are provided by bilinear pairing, but real-world testing and scalability in diverse healthcare setups could be further explored. | [143] |
| | | An XACML-oriented framework, implementing a prototype known as proof-of-concept and analysis. | The policy-based approach neatly analyzed access to services for authorized users while protecting important data from intrusion. | Access control in constrained healthcare environments | This lowered the policy number needed for user access control specification settings. | [123] |
| | Secure architecture | A network forensics framework implementing Deep Learning. | Identification and tracing of attacking behaviors within IoT networks. | Tracing of events of cyber-attacks. | Results implied a high-performing system for discovering and tracing cyber-attack events. | [132] |
| | | Introduce a lightweight deep learning security framework for real-time | Enhanced intrusion detection accuracy using a hybrid MobileNetV3-SVM model with reduced computational overhead. | Real-time threat detection in IoT and 5G edge networks, ensuring secure | Efficient and adaptable, but further optimization for large-scale deployment and | [154] |

| | | intrusion detection in IoT and 5G networks. | | communication and adaptive risk mitigation. | evaluation of evolving threat landscapes may be needed. | |
| --- | --- | --- | --- | --- | --- | --- |
| | | A blockchain-based shared audit framework (BSAF) for analyzing digital forensics data. | The framework was created to detect and trace the causative agent of data scavenging attacks in VR (virtualized resources). Blockchain technology was used for accessing logs and managing control. | Data scavenging attacks in VR. | Results showed that the framework is consistent and reduces analysis time and adversary event rate. | [134] |

Various forms of attacks, aimed specifically at IoT devices and networks, have emerged as a direct result of the proliferation of IoT [158]. Some of the common IoT attacks include denial of service attacks, distributed DoS, jamming attacks, intrusion detection system attacks, botnet attacks, malicious node attacks, power analysis attacks, side channel security attacks, device exploitation, man-in-the-middle attacks, physical attacks, wormhole attacks, Sybil attacks, zero-day exploits, routing attacks, buffer overflow attacks, supply chain attacks, impersonation attacks, eavesdropping and data interception, firmware tampering, and spoofing attacks. These common attacks are summarized in Table 6. Security techniques such as encryption protocols, strong authentication procedures, network segmentation, regular firmware updates, and monitoring for aberrant behavior are essential for protecting against these IoT attacks.

Table 6. Common attacks in IoT, along with their description

| Attacks | Description | Ref. |
| --- | --- | --- |
| Denial of service (DoS) attacks | DoS attacks, which consist of a mass of network packets aimed at a specific application node, disrupt service in real time. This attack is frequently seen in the IoT applications. | [159] |
| Distributed DoS | Distributed DoS attacks include making a server unavailable and preventing network nodes with intelligence from accessing the resources they need. | [160] |
| Jamming attacks | A type of DoS attack in which the attacker attempts to disrupt communication. | [161] |
| Intrusion detection system (IDS) | An IDS monitors network activity for unauthorized changes. Misuse detection, host-based IDS, network-based IDS, and anomaly detection are different kinds of IDS attacks. | [162] |
| Botnet attacks | Insecure IoT devices are prime targets for employment in botnets like the infamous Mirai botnet. Distributed DoS attacks, launched from these botnets, can overwhelm their targets and disrupt service. | [163] |
| Malicious node attacks | Due to the diverse nature of IoT devices, malicious node attacks are a real threat in decentralized networks. The malicious nodes can be identified, and trust between the nodes can be built using perception and K-means. | [164] |

| Power analysis attacks | Power analysis attacks mainly occur to increase the computing capacity of the nodes to the point where the fundamental cryptographic algorithm cannot be executed. To establish node-to-node trust in an IoT network, privacy must be maintained. | [165] |
|---|---|---|
| Device exploitation | Attackers exploit devices to obtain access to or control IoT infrastructure. This may involve taking advantage of weak or default passwords, unpatched hardware/software, or unsecured methods for communication. Once infiltrated, attackers can use devices to perform unauthorized actions, steal information, or gain access to other networks. | [166] |
| Man-in-the-middle attacks | Attackers can access sensitive information or modify existing files by intercepting communications between IoT devices and backend services. Rogue access points and corrupted network infrastructure are two common causes of this problem. | [167] |
| Physical attacks | Attacks, including device theft, hardware tampering, and side-channel attacks, are all made possible by having physical access to IoT devices. A device's security can be compromised in a number of ways, including the theft of encryption keys, the manipulation of sensors, and the injection of malicious components. | [168] |
| Wormhole attacks | At the 6LoWPAN layer, an attacker can launch a wormhole attack by creating a tunnel between two connected nodes. | [169] |
| Sybil attacks | Sybil Attacks are cybercrimes in which an individual creates several fake accounts on a distributed database to breach security. | [170] |
| Zero-day exploits | Zero-day exploits are used by attackers to take advantage of vulnerabilities in IoT devices before the manufacturers patch them. These vulnerabilities allow attackers to take control of systems or launch additional attacks. | [171] |
| Buffer overflow attacks | The Buffer overflow exploit is the act of writing a program into an insufficiently sized memory block. The IoT network is susceptible to a memory overflow attack while nodes execute various programs on devices for processing or computing purposes. | [172] |

| Supply chain attacks | Because of the complex structure of the IoT supply chain, there are several points of vulnerability in the hardware, software, and distribution channels where devices can be compromised. Devices can be infiltrated before they ever reach end users by inserting malicious firmware or hardware implants. | [173] |
|---|---|---|
| Impersonation attacks | An impersonation attack is one in which a malicious node poses as a trustworthy one to steal data from other nodes in the network. Smart devices in IoT applications tend to be low-cost and diverse, making this a difficult problem to solve. | [174] |
| Eavesdropping and data interception | Sensitive data sent between IoT devices might be compromised by insecure communication pathways and inadequate encryption. Attackers can eavesdrop on communications and potentially steal sensitive information or violate users' privacy rights. | [175] |
| Firmware tampering | Attackers can modify the firmware of IoT devices to change their behavior or insert malicious code. This can make the device vulnerable to attacks in which it is used as part of a broader network of compromised machines. | [176] |
| Spoofing attacks | Spoofing attacks entail unauthorized access to private data stored on IoT-connected devices. Spoofing attacks can benefit from this data in the future. | [177] |

In conclusion, insufficient encryption or poorly deployed network protocols might compromise data confidentiality and integrity. In addition, devices are susceptible to known exploits and vulnerabilities due to the lack of frequent firmware upgrades and security patches from manufacturers. Concerns about privacy arise because of the large amounts of personal and sensitive data acquired by IoT devices, which could be disclosed or misused if suitable data handling policies and consent mechanisms are not in place. Another concern is the lack of protection against physical theft or tampering with IoT gadgets used in the real world. Distributed denial of service attacks that leverage compromised IoT devices are becoming more common. Providing consistent and effective security measures across the IoT ecosystem is also complicated by the lack of unified security standards and compatibility among IoT devices and platforms.

Strong authentication systems, encryption protocols, regular security updates, privacy measures, and industry-wide security standards are all necessary to address these security challenges, and they require a multi-pronged strategy involving manufacturers, developers, users, and regulatory agencies.

## 5. Research challenges in IoT

IoT intrigues researchers and scientists owing to its vast scale and scope. In recent decades, scientists have put significant efforts into various disciplines of IoT, including its architecture security, application development, reliability, protocols, privacy, and connectivity. Many enterprises are engaged in IoT application projects and progress in several sectors. For instance, major international standards development organizations (SDOs), such as ISO, IEC, IETF, ITU, and IEEE, are scrutinizing the need for the success of IoT [178].

### 5.1 Technical challenges

Although technologies like blockchain, fog, edge computing, and low power wide area network (LPWAN) work to enhance or protect IoT activities, implementing them has some challenges. For instance, blockchain technology adds a substratum of protection to accumulate IoT data that hackers must bypass to ingress the system. It provides relatively higher encryption standards, making it highly unfeasible to alter existing data entries [179,180]. However, the current blockchain infrastructure is inadequate for humongous IoT networks due to the limited node numbers in public blockchain networks and the limited latency in permission-less networking.

The development of consensus methods to limit redundant processing power utilization in resource-restrained devices remains a challenge in this field of research [181]. Furthermore, the inception of trash data is a crucial element of blockchain's tamper-proof aspect as it influences the ultimate execution of the underlying program. Consequently, it entails suitable garbage data managing procedures. Service providers and manufacturers also confront regulatory and legal challenges, which hinder the acquisition of blockchain technology in the workplace. Another significant stumbling block is storage, as the ledger must be stored on the nodes but on a central server. The inability to comprehend and derive the potential of blockchain in the IoT area harms its appeal [182].

Ensuring robust data security, compliance with regulatory standards, ethical alignment with societal values, and the reliability and accuracy of IoT devices is essential for safeguarding users' safety [183]. Excessive reliance on IoT technology makes supply chain management vulnerable to technological dependencies and single points of failure [184]. Failures or disruptions in IoT devices, communication networks, or cloud infrastructure can lead to widespread operational interruptions and substantial financial losses.

Devices, such as sensors, embedded devices, RFID tags, etc., with scarce resources at the network's edge, are prone to a range of malicious activities. Because these devices are primarily reliant on external power sources, blackouts can halt the entire edge layer [185]. Furthermore, deploying security protocols on these devices adds to their computational overhead. Subsequently, LPWAN aims to reach optimal IoT network performance. However, tackling interference challenges in LPWAN (co-layer, cross-layer, and inter-symbol) is still a continuing process. Unauthorized technologies similar to LoRa and SigFox are one cause of these problems, precisely due to their operator ISM bands that cause the interruption [186].

Due to the exponential decline in the LPWANs' performance, ensuring scalability in heavy IoT networks is another crucial concern. Effective energy handling in IoT devices also presents an existing issue that needs to be remedied. Low-power communications is an essential research discipline because battery renewal is expensive, particularly for vast implementations, such as IoT. While many energy-efficient strategies exist at multiple-element layers of the Open System Interconnection (ISO) paradigm, batteries must be substituted occasionally, which is a prodigious hurdle to adopting IoT technology [187]. Scalability, one of the most significant technological issues, arises from the large interlinked objects connected with IoT. The primary intent of creating a scalable gadget is to fulfill shifting needs that can never be static because people's interests and tastes change over time and under different circumstances. All the devices are expected to be accommodated in the address space.

To retrieve and save relevant information from enormous amounts of data, adequate information extraction procedures must be applied. In IoT, service discovery is challenging because we are expected to find the node that confers the client request among interconnected devices [187]. To solve the challenges surrounding blockchain, fog, and edge computing in the future, guidelines must be formulated globally to maintain optimal IoT execution. The United States, Malta, Gibraltar, and Belarus are some countries working on these guidelines.

Implementing end-to-end networks in fog computing would lead to more efficient trade-offs between distributed and centralized network configurations. Adopting artificial intelligence and machine learning approaches can help develop intelligent fog nodes that determine adaptive data management and relaying decisions [187]. Meanwhile, Named Data Networking (NDN) is a revolutionary naming scheme for edge computing that ensures effective scalability [188,189].

**5.2 Security assurance and privacy concerns**

IoT demands the appropriate address of security and trust functions since it has become a critical component of the growth of the Internet with its massive utilization. Core IoT networks need authorization, authentication, confidentiality, non-repudiation, and integrity to be considered secure [190]. Many cyberattacks exploit vulnerabilities in specific devices to access their networks, leaving protected appliances vulnerable [191,192]. Researchers are conscious of the existing flaws because IoT is built on top of conventional wireless sensor networks (WSNs), thus inheriting the same protection concerns that WSN faces [193,194]. As most of the nodes utilized in IoT systems have memory, energy, and other limits, adopting robust security approaches is a complex phenomenon that needs to be appropriately handled by researchers.

The four dominant categories of cyberattacks affecting the IoT system are software, network, encryption, and physical [195]. Because IoT extensively uses a wireless interface for data transmission, message forging, interference, and eavesdropping present some threats. This privacy vacuum further stimulates efficient security alternatives, consisting of probes in non-cryptographic security mechanisms and frameworks that assist in developing strong encryption on heterogeneous systems. Despite various anonymization strategies proposed for ensuring security and privacy, most of them require a massive number of resources, such as power, memory, and bandwidth in actual use. As a result, traditional IP-based security protocols are insufficient in IoT networks [187,196].

Extensive procedures and regulations must be established to guide the disclosure of personal data alongside technological improvements to ensure that unscrupulous parties do not infiltrate such information. Several studies have developed algorithms and frameworks to tackle privacy and security challenges. For instance, Pacheco and Hariri [197] recommended a security framework that provides a secure system in smart infrastructure (i.e., smart buildings) and proffers a reliable system. The proposed network has IoT service, node, network, and application layers.

The researchers also established a threat model to discern the vulnerabilities in each layer and the plausible countermeasures required to prevent their swindling. The model can identify IoT nodes and the suitable mitigation method instead of encrypting the data owing to the truncated computing potency of IoT devices. Although it is possible to reuse some conventional IP-based security mechanisms, the limited resource features must be regarded. As a result, hardware security support and numerous IP-based safety protocol versions, such as Datagram HIP Diet Exchange, TLS, lightweight encryption, and minimal IKEv2, have recently been created [146,198–200]. However, effective security strategies are necessary for the future, precisely for data transmission and software updates, as conventional security techniques are expensive to deploy.

## 5.3 Standardization issues

Communication protocols, application requirements, object identification, security, information processing, data, and service platforms require standardization [198,201]. It enables interoperability, promoting seamless information exchange and integration among distributed applications. The rise in critical cybersecurity threats has also heightened the demand for advanced IoT infrastructure configurations [202]. This necessitates standardized platforms and protocols for communication across multiple devices and vendors. Unifying IoT standards could also enhance general security, as linking equipment from distinct vendors would be facile to secure [203]. LPWPANs in place of IPv6, Routing over Power and Lossy Networks, and Constricted Restful Environments are intended to expand web design to the most constrained networks and embedded systems [204]. Although the ETSI and IETF institutions are working on developing IoT standards, the universal communication standard remains a research challenge in IoT [205].

Standardized IoT communication technologies and architecture are the future cornerstones of IoT evolution. W3C, ITU, ETSI, IEEE, NIST, IETF, OneM2M, OASIS, and other leading entities are collaborating to provide a standard IoT framework. Nevertheless, integrating distinct criteria and initiatives in a cohesive sense is challenging as it may lead to inadequate interfaces, incompleteness, ambiguity, and poor maintenance [206]. Additionally, several experts are developing a unified architecture for IoT, which requires compliance across existing schemes and tackles heterogeneity abstraction and interoperability, etc. Data must be emitted to other devices in IoT using effective routing protocols. Nevertheless, implementing effective communication conventions is challenging because of IoT and WSN features like scarce resources and Lossy wireless networks [187]. The potential solutions associated with the challenges in IoT are summarized in Table 7.

Table 7. Challenges in IoT and their potential solutions

| Challenges | Potential solutions | Remarks | Ref. |
|---|---|---|---|
| Technical | Implementing End-to-end networks | Provide fog computing with a more efficient configuration | [187] |
| | Adopting AI and ML | Develop intelligent fog nodes that determine adaptive data management. | |
| | NDN framework | Ensure effective scalability | [188,189] |
| Security & Privacy | Security Framework | Essential to develop security features, compatible hardware, and third-party smart apps | [146] |
| | Light-weight block cipher | Expanding the applications of cryptography to constrained devices and its related international guidelines and standardization compilation are recently underway. | [200] |
| Standardization | Internet Engineering Task Force (IETF) | Connect the limited nodes in a restricted environment efficiently | [207] |
| | IoT Azure Hub | Support a wide range of environments, standards, processing patterns, and scenarios with high scalability and security. | [203] |
| | SmartThings | It can connect several goods and functions as a complete home hub, compatible with various standards for various intelligent systems. However, the hub connection is relatively slow | |

Overall, there is a wide variety of noticeable challenges that researchers in the IoT field must solve to ensure the reliable and secure operation of connected devices and systems. Major obstacles involve dealing with the scalability and difficulty of IoT deployments, ensuring security and privacy in the face of rising cyber threats, developing interoperability and standardizing among different IoT platforms, maximizing energy efficiency in limited resource environments, making efficient data analytics and decision-making techniques, taking on the challenges of fog and edge computing architectures, and dealing with the social and ethical implications of IoT. To overcome these obstacles and guide the IoT toward a more responsible and sustainable future, we need to invest heavily in collaboration, research, and innovation.

## 6. Future prospects of IoT

Recently, the future of IoT has been a key field of study, allowing various actuators, IoT sensors, and diverse objects to communicate specifically without human interaction. Furthermore, with critical cybersecurity threats, the demand and standards for a sophisticated IoT infrastructure are rapidly increasing [187]. Several next-generation configurations (i.e. Named Data Networking [NDN]) can be used to address and name features in IoT applications. NDN can also be operated in IoT instead of IPv4 or IPv6 addresses, and incorporating NDN in IoT infrastructure has already been investigated by some experts [208]. Further research is conceivable because the Internet is heading towards a content-centric paradigm [5].

Seeking novel IoT applications in Content-Centric Networking (CCN) and DTN is another step toward a more futuristic vision. Guo et al. [209] offered the first effort to build a delay-tolerant variant of IoT by linking gadgets to generate infrastructure-less networks leveraging short-term communication systems, including WiFi or Bluetooth. Data dissemination and pooling across devices in such an opportunistic IoT are initiated based on mobility and fortuitous interactions. Since P2P CNN techniques are becoming more prevalent in Internet applications, integrating IoT into this context is another prospect to inspect [210]. Similarly, only a few studies have explored IoT underwater and how to develop delay-tolerant IoT systems, both of which are attractive themes for continued investigations in the future [211,212].

Fifth-generation (5G) cellular networks are crucial determinants for the widespread acceptance of IoT technology. Massive connectivity, extensive coverage, high security and privacy, ultra-low latency, and ultra-high reliability are imperative in new business models intended for IoT deployment. Furthermore, effective security strategies are necessary for the future, precisely for data transmission and software updates, as conventional security techniques are expensive to deploy. Data volume rates, wider coverage, and high bandwidth are features of the renowned 5G-enabled IoT that answer business models and allow IoT to be used for actuators, drones, and robotics [193]. Future intelligent platforms require IoT nodes to work in various operational environments and with the cloud and network to enhance system intelligence while conserving energy consumption.

Many practical applications demand data analysis on vast scales. As a result, future researchers may confront intense challenges in producing Big Data analysis for prospective IoT systems. To sum up, the prospects for IoT in the future are boundless. IoT's potency is in activating billions of devices concurrently and harnessing massive usable data, streamlining a wide range of business operations. An exhilarating wave of IoT applications will emerge when IoT platforms overcome the challenges by expanding efficiency and artificial intelligence. This will necessitate strong collaboration between IoT and network platforms to introduce IoT in the future. Future intelligent platforms require IoT nodes to work in various operational environments and with the cloud and network to enhance system intelligence while conserving energy consumption [213].

In summary, the IoT is predicted to grow in scope as technology develops, leading to an explosion of wirelessly networked devices that penetrate every aspect of modern life. New opportunities for real-time use and data-intensive IoT systems will be made possible with the advent of 5G networks, allowing quicker and more dependable communication. More responsive and efficient IoT deployments will be achievable as edge computing develops, allowing for localized data processing and lowering latency. Automation, predictive analytics, and individualized user experiences are all potentially feasible by the massive amounts of data produced by IoT devices. AI and ML will play a crucial role in this process. Developments in security and privacy controls will be essential to maintain reliability and protect private data in the IoT ecosystem. Healthcare, transportation, manufacturing, and agriculture are some sectors that will undergo radical change due to IoT's capacity to drive efficiency, innovation, and sustainability. With an ever-expanding network of linked devices, transformative- applications, and constant connection, IoT's future is promising.

## 7. Conclusions

IoT can collect and share data with other objects and systems over the internet or any other network. However, there is a dearth of discussion and a lack of sufficiently detailed IoT literature to address the identified challenges. This review, thus, examined the architecture, current progress, security and forensic challenges, applications, and research challenges of IoT by analyzing and reviewing many new and emerging techniques and developments. The findings demonstrate that it is pertinent to develop a high-performing framework for network forensics that can detect and trace cyberattack incidents using deep learning. Additional security can be provided by utilizing a blockchain-based authentication system with a multi-chain structure connecting several trust domains. In a Fog-based IoT forensic framework, additional encryption makes it more challenging to share data. As a solution, a FLIP-based system can be developed using IoT, which is very scalable and adaptive to meet consumers' security-conscious expectations. Therefore, researchers who use big data to build IoT-based smart systems should carefully contemplate how much data is created and supplied when designing such systems. Strong protections are also needed to keep IoT systems safe from cyber attacks, as privacy and security issues remain crucial.

IoT's wide range of applications shows its revolutionary potential across industries, from healthcare monitoring and environmental management to smart homes and industrial automation. Interoperability, scalability, data analytics, energy efficiency, and ethical concerns are still major obstacles to study. Despite these difficulties, IoT's future looks promising. The introduction of 5G networks, developments in edge computing, AI, and improved standards will all contribute to the IoT expansion. Incorporating IoT into society will transform industries, stimulate innovation, and increase productivity. To realize the full potential of IoT as it develops, it is essential to take on these difficulties and promote sustainable and responsible IoT installations.

**Data availability**

No data were used in the research presented in the article.

**Conflict of interest**

No authors have disclosed any conflicts of interest that would have been perceived as having influenced the work presented in this paper, either financially or otherwise.

**References**

[1] K.C. Rath, A. Khang, D. Roy, The Role of Internet of Things (IoT) Technology in Industry 4.0 Economy, in: Adv. IoT Technol. Appl. Ind. 4.0 Digit. Econ., 2024. https://doi.org/10.1201/9781003434269-1.

[2] X. Mu, M.F. Antwi-Afari, The applications of Internet of Things (IoT) in industrial management: a science mapping review, Int. J. Prod. Res. 62 (2024). https://doi.org/10.1080/00207543.2023.2290229.

[3] J.H. Nord, A. Koohang, J. Paliszkiewicz, The Internet of Things: Review and theoretical framework, Expert Syst. Appl. 133 (2019) 97–108. https://doi.org/10.1016/j.eswa.2019.05.014.

[4] M. Nassereddine, A. Khang, Applications of Internet of Things (IoT) in Smart Cities, in: Adv. IoT Technol. Appl. Ind. 4.0 Digit. Econ., 2024. https://doi.org/10.1201/9781003434269-6.

[5] C.C. Sobin, A Survey on Architecture, Protocols and Challenges in IoT, Wirel. Pers. Commun. 112 (2020). https://doi.org/10.1007/s11277-020-07108-5.

[6] A.L. Duguma, X. Bai, Contribution of Internet of Things (IoT) in improving agricultural systems, Int. J. Environ. Sci. Technol. (2023). https://doi.org/10.1007/s13762-023-05162-7.

[7] S.F. Ahmed, M.S. Bin Alam, S. Afrin, S.J. Rafa, N. Rafa, A.H. Gandomi, Insights into Internet of Medical Things (IoMT): Data fusion, security issues and potential solutions, Inf. Fusion 102 (2024). https://doi.org/10.1016/j.inffus.2023.102060.

[8] J.M. Talavera, L.E. Tobón, J.A. Gómez, M.A. Culman, J.M. Aranda, D.T. Parra, L.A. Quiroz, A. Hoyos, L.E. Garreta, Review of IoT applications in agro-industrial and environmental fields, Comput. Electron. Agric. 142 (2017) 283–297. https://doi.org/10.1016/J.COMPAG.2017.09.015.

[9] L. Miao, K. Liu, Poster Abstract: Towards a heterogeneous internet-of-Things testbed via mesh inside a mesh, Proc. 14th ACM Conf. Embed. Networked Sens. Syst. SenSys 2016 (2016) 368–369. https://doi.org/10.1145/2994551.2996710.

[10] A.T. Rosário, J.C. Dias, Exploring the Landscape of Smart Tourism: A Systematic Bibliometric Review of the Literature of the Internet of Things, Adm. Sci. 14 (2024). https://doi.org/10.3390/admsci14020022.

[11] P.P. Ray, A survey on Internet of Things architectures, J. King Saud Univ. - Comput. Inf. Sci. 30 (2018) 291–319. https://doi.org/10.1016/J.JKSUCI.2016.10.003.

[12] N. Sathiyanatha, P. Selvaprasanth, A Brief Study on IoT Applications, Int. J. Trend Sci. Res. Dev. 4 (2020) 2456–6470.

[13] M. Ghobaei-Arani, A.A. Rahmanian, A. Souri, A.M. Rahmani, A moth-flame optimization algorithm for web service composition in cloud computing: Simulation and verification, Softw. - Pract. Exp. 48 (2018) 1865–1892. https://doi.org/10.1002/spe.2598.

[14] S.F. Ahmed, M.S. Bin Alam, S. Afrin, S.J. Rafa, S.B. Taher, M. Kabir, S.M. Muyeen, A.H. Gandomi, Toward a Secure 5G-Enabled Internet of Things: A Survey on Requirements, Privacy, Security, Challenges, and Opportunities, IEEE Access 12 (2024). https://doi.org/10.1109/ACCESS.2024.3352508.

[15] M. Casillo, L. Cecere, F. Colace, A. Lorusso, D. Santaniello, Integrating the Internet of Things (IoT) in SPA Medicine: Innovations and Challenges in Digital Wellness, Computers 13 (2024). https://doi.org/10.3390/computers13030067.

[16] M. Conti, A. Dehghantanha, K. Franke, S. Watson, Internet of Things security and forensics: Challenges and opportunities, Futur. Gener. Comput. Syst. 78 (2018) 544–546. https://doi.org/10.1016/j.future.2017.07.060.

[17] E. Bertino, K.K.R. Choo, D. Georgakopolous, S. Nepal, Internet of things (IoT): Smart and secure service delivery, ACM Trans. Internet Technol. 16 (2016) 1–7. https://doi.org/10.1145/3013520.

[18] S.R. Moosavi, T.N. Gia, E. Nigussie, A.M. Rahmani, S. Virtanen, H. Tenhunen, J. Isoaho, End-to-end security scheme for mobility enabled healthcare Internet of Things, Futur. Gener. Comput. Syst. 64 (2016) 108–124. https://doi.org/10.1016/j.future.2016.02.020.

[19] N. Xiao, Z. Wang, X. Sun, J. Miao, A novel blockchain-based digital forensics framework for preserving evidence and enabling investigation in industrial Internet of Things, Alexandria Eng. J. 86 (2024). https://doi.org/10.1016/j.aej.2023.12.021.

[20] Sakshi, A. Malik, A.K. Sharma, A survey on blockchain based IoT forensic evidence preservation: research trends and current challenges, Multimed. Tools Appl. 83 (2024). https://doi.org/10.1007/s11042-023-17104-z.

[21] G. Kumar, R. Saha, C. Lal, M. Conti, Internet-of-Forensic (IoF): A blockchain based digital forensics framework for IoT applications, Futur. Gener. Comput. Syst. 120 (2021)

13–25. https://doi.org/10.1016/j.future.2021.02.016.

[22] S. Brotsis, N. Kolokotronis, K. Limniotis, S. Shiaeles, D. Kavallieros, E. Bellini, C. Pavue, Blockchain solutions for forensic evidence preservation in iot environments, Proc. 2019 IEEE Conf. Netw. Softwarization Unleashing Power Netw. Softwarization, NetSoft 2019 (2019) 110–114. https://doi.org/10.1109/NETSOFT.2019.8806675.

[23] Shruti, S. Rani, W. Boulila, Securing internet of things device data: An ABE approach using fog computing and generative AI, Expert Syst. (2024). https://doi.org/10.1111/exsy.13691.

[24] E. Al-Masri, Y. Bai, J. Li, A fog-based digital forensics investigation framework for IoT systems, in: Proc. - 3rd IEEE Int. Conf. Smart Cloud, SmartCloud 2018, Institute of Electrical and Electronics Engineers Inc., 2018: pp. 196–201. https://doi.org/10.1109/SmartCloud.2018.00040.

[25] H. Mahmood, M. Arshad, I. Ahmed, S. Fatima, H. ur Rehman, Comparative study of IoT forensic frameworks, Forensic Sci. Int. Digit. Investig. 49 (2024). https://doi.org/10.1016/j.fsidi.2024.301748.

[26] M. Harbawi, A. Varol, An improved digital evidence acquisition model for the Internet of Things forensic I: A theoretical framework, 2017 5th Int. Symp. Digit. Forensic Secur. ISDFS 2017 (2017). https://doi.org/10.1109/ISDFS.2017.7916508.

[27] C.J. D'Orazio, K.K.R. Choo, L.T. Yang, Data Exfiltration from Internet of Things Devices: IOS Devices as Case Studies, IEEE Internet Things J. 4 (2017) 524–535. https://doi.org/10.1109/JIOT.2016.2569094.

[28] S. Watson, A. Dehghantanha, Digital forensics: the missing piece of the Internet of Things promise, Comput. Fraud Secur. 2016 (2016) 5–8. https://doi.org/10.1016/S1361-3723(15)30045-2.

[29] N. Mishra, S. Pandya, Internet of Things Applications, Security Challenges, Attacks, Intrusion Detection, and Future Visions: A Systematic Review, IEEE Access 9 (2021) 59353–59377. https://doi.org/10.1109/ACCESS.2021.3073408.

[30] K.M. Sadique, R. Rahmani, P. Johannesson, Towards security on internet of things: Applications and challenges in technology, Procedia Comput. Sci. 141 (2018) 199–206. https://doi.org/10.1016/j.procs.2018.10.168.

[31] M.K. Saini, R.K. Saini, Internet of Things (IoT) Applications and Security Challenges: A

Review, Int. J. Eng. Res. Technol. 7 (2019) 1–7.

[32] A. Raghuvanshi, U. Kumar Singh, M. Shuaib, S. Alam, An investigation of various applications and related security challenges of Internet of things, Mater. Today Proc. (2021). https://doi.org/10.1016/j.matpr.2021.01.821.

[33] H.U. Rehman, M. Asif, M. Ahmad, Future applications and research challenges of IOT, 2017 Int. Conf. Inf. Commun. Technol. ICICT 2017 2017-Decem (2018) 68–74. https://doi.org/10.1109/ICICT.2017.8320166.

[34] M. Stoyanova, Y. Nikoloudakis, S. Panagiotakis, E. Pallis, E.K. Markakis, A Survey on the Internet of Things (IoT) Forensics: Challenges, Approaches, and Open Issues, IEEE Commun. Surv. Tutorials 22 (2020) 1191–1221. https://doi.org/10.1109/COMST.2019.2962586.

[35] H.F. Atlam, E.E.-D. Hemdan, A. Alenezi, M.O. Alassafi, G.B. Wills, Internet of things forensics: A review, Internet of Things 11 (2020) 100220.

[36] I. Yaqoob, I.A.T. Hashem, A. Ahmed, S.M.A. Kazmi, C.S. Hong, Internet of things forensics: Recent advances, taxonomy, requirements, and open challenges, Futur. Gener. Comput. Syst. 92 (2019) 265–275.

[37] C. Mahmoud, S. Aouag, Security for internet of things: A state of the art on existing protocols and open research issues, in: Proc. 9th Int. Conf. Inf. Syst. Technol., 2019: pp. 1–6.

[38] M.M. Ogonji, G. Okeyo, J.M. Wafula, A survey on privacy and security of Internet of Things, Comput. Sci. Rev. 38 (2020) 100312. https://doi.org/https://doi.org/10.1016/j.cosrev.2020.100312.

[39] S.M. Tahsien, H. Karimipour, P. Spachos, Machine learning based solutions for security of Internet of Things (IoT): A survey, J. Netw. Comput. Appl. 161 (2020) 102630. https://doi.org/https://doi.org/10.1016/j.jnca.2020.102630.

[40] M. Aly, F. Khomh, M. Haoues, A. Quintero, S. Yacout, Enforcing security in Internet of Things frameworks: A Systematic Literature Review, Internet of Things 6 (2019) 100050. https://doi.org/https://doi.org/10.1016/j.iot.2019.100050.

[41] A. Azmoodeh, A. Dehghantanha, K.-K.R. Choo, Big data and internet of things security and forensics: Challenges and opportunities, Handb. Big Data IoT Secur. (2019) 1–4.

[42] W. Yang, M.N. Johnstone, L.F. Sikos, S. Wang, Security and forensics in the internet of

things: research advances and challenges, in: 2020 Work. Emerg. Technol. Secur. IoT, IEEE, 2020: pp. 12–17.

[43] S.U. Qureshi, J. He, S. Tunio, N. Zhu, A. Nazir, A. Wajahat, F. Ullah, A. Wadud, Systematic review of deep learning solutions for malware detection and forensic analysis in IoT, J. King Saud Univ. - Comput. Inf. Sci. 36 (2024). https://doi.org/10.1016/j.jksuci.2024.102164.

[44] G. Chen, T. Jiang, M. Wang, X. Tang, W. Ji, Design and model checking of timed automata oriented architecture for Internet of thing, Int. J. Distrib. Sens. Networks 16 (2020). https://doi.org/10.1177/1550147720911008.

[45] R. Salama, F. Al-Turjman, M. Aeri, S.P. Yadav, Internet of Intelligent Things (IoT) - An Overview, in: 2023 Int. Conf. Comput. Intell. Commun. Technol. Networking, CICTN 2023, 2023. https://doi.org/10.1109/CICTN57981.2023.10141157.

[46] W. Razouk, D. Sgandurra, K. Sakurai, A new security middleware architecture based on fog computing and cloud to support IoT constrained devices, in: ACM Int. Conf. Proceeding Ser., Association for Computing Machinery, 2017. https://doi.org/10.1145/3109761.3158413.

[47] J.M. Kizza, Internet of Things (IoT): Growth, Challenges, and Security, in: 2024. https://doi.org/10.1007/978-3-031-47549-8_25.

[48] D.P. Abreu, K. Velasquez, M. Curado, E. Monteiro, A resilient Internet of Things architecture for smart cities, Ann. Des Telecommun. Telecommun. 72 (2017) 19–30. https://doi.org/10.1007/s12243-016-0530-y.

[49] U. Javaid, B. Sikdar, A Checkpoint Enabled Scalable Blockchain Architecture for Industrial Internet of Things, IEEE Trans. Ind. Informatics 17 (2021) 7679–7687. https://doi.org/10.1109/TII.2020.3032607.

[50] H. Mrabet, S. Belguith, A. Alhomoud, A. Jemai, A survey of IoT security based on a layered architecture of sensing and data analysis, Sensors (Switzerland) 20 (2020) 1–20. https://doi.org/10.3390/s20133625.

[51] T.A.A.A.A. Bary, B.M. Elomda, H.A. Hassan, Multiple Layer Public Blockchain Approach for Internet of Things (IoT) Systems, IEEE Access 12 (2024) 56431–56438. https://doi.org/10.1109/ACCESS.2024.3389299.

[52] S. Hartono, R. Kosala, S.H. Supangkat, B. Ranti, Smart Hybrid Learning Framework

Based on Three-Layer Architecture to Bolster Up Education 4.0, Proceeding - 2018 Int. Conf. ICT Smart Soc. Innov. Towar. Smart Soc. Soc. 5.0, ICISS 2018 (2018) 1–5. https://doi.org/10.1109/ICTSS.2018.8550028.

[53] K. Kethineni, G. Pradeepini, Intrusion detection in internet of things-based smart farming using hybrid deep learning framework, Cluster Comput. 27 (2024). https://doi.org/10.1007/s10586-023-04052-4.

[54] M. Parto, C. Saldana, T. Kurfess, A novel three-layer IoT architecture for shared, private, scalable, and real-time machine learning from ubiquitous cyber-physical systems, in: Elsevier B.V., 2020: pp. 959–967. https://doi.org/10.1016/j.promfg.2020.05.135.

[55] J.Y. Zhu, B. Tang, V.O.K. Li, A five-layer architecture for big data processing and analytics A five-layer architecture for big data processing and analytics 39, 2019.

[56] L. Bittencourt, R. Immich, R. Sakellariou, N. Fonseca, E. Madeira, M. Curado, L. Villas, L. DaSilva, C. Lee, O. Rana, The Internet of Things, Fog and Cloud continuum: Integration and challenges, Internet of Things 3–4 (2018) 134–155. https://doi.org/10.1016/J.IOT.2018.09.005.

[57] A.A. Al-Atawi, Enhancing data management and real-time decision making with IoT, cloud, and fog computing, IET Wirel. Sens. Syst. (2024). https://doi.org/10.1049/wss2.12099.

[58] A.P. Plageras, K.E. Psannis, C. Stergiou, H. Wang, B.B. Gupta, Efficient IoT-based sensor BIG Data collection–processing and analysis in smart buildings, Futur. Gener. Comput. Syst. 82 (2018) 349–357.

[59] S. Dhingra, R.B. Madda, R. Patan, P. Jiao, K. Barri, A.H. Alavi, Internet of things-based fog and cloud computing technology for smart traffic monitoring, Internet of Things 14 (2021) 100175.

[60] E. Huaranga-Junco, S. González-Gerpe, M. Castillo-Cara, A. Cimmino, R. García-Castro, From cloud and fog computing to federated-fog computing: A comparative analysis of computational resources in real-time IoT applications based on semantic interoperability, Futur. Gener. Comput. Syst. 159 (2024) 134–150. https://doi.org/10.1016/j.future.2024.05.001.

[61] Q.D. La, M. V Ngo, T.Q. Dinh, T.Q.S. Quek, H. Shin, Enabling intelligence in fog computing to achieve energy and latency reduction, Digit. Commun. Networks 5 (2019)

3–9.

[62] E. Gomes, F. Costa, C. De Rolt, P. Plentz, M. Dantas, A survey from real-time to near real-time applications in fog computing environments, in: Telecom, MDPI, 2021: pp. 489–517.

[63] Shruti, S. Rani, G. Srivastava, Secure hierarchical fog computing-based architecture for industry 5.0 using an attribute-based encryption scheme, Expert Syst. Appl. 235 (2024). https://doi.org/10.1016/j.eswa.2023.121180.

[64] V. Gupta, H. Singh Gill, P. Singh, R. Kaur, An energy efficient fog-cloud based architecture for healthcare, J. Stat. Manag. Syst. 21 (2018) 529–537. https://doi.org/10.1080/09720510.2018.1466961.

[65] M.R. Alizadeh, V. Khajehvand, A.M. Rahmani, E. Akbari, Task scheduling approaches in fog computing: A systematic review, Int. J. Commun. Syst. 33 (2020) e4583.

[66] A. Javadpour, F. Ja'fari, T. Taleb, C. Benzaid, Y. Bin, Y. Zhao, Encryption as a Service (EaaS): Introducing the Full-Cloud-Fog Architecture for Enhanced Performance and Security, IEEE Internet Things J. (2024). https://doi.org/10.1109/JIOT.2024.3450192.

[67] D.H. Abdulazeez, S.K. Askar, A Novel Offloading Mechanism Leveraging Fuzzy Logic and Deep Reinforcement Learning to Improve IoT Application Performance in a Three-Layer Architecture Within the Fog-Cloud Environment, IEEE Access 12 (2024). https://doi.org/10.1109/ACCESS.2024.3376670.

[68] H. Li, Z. Gu, L. Deng, Y. Han, C. Yang, Z. Tian, A fine-grained video encryption service based on the cloud-fog-local architecture for public and private videos, Sensors (Switzerland) 19 (2019). https://doi.org/10.3390/s19245366.

[69] J. Yue, Z. Hu, R. He, X. Zhang, J. Dulout, C. Li, J.M. Guerrero, Cloud-fog architecture based energy management and decision-making for next-generation distribution network with prosumers and internet of things devices, Appl. Sci. 9 (2019). https://doi.org/10.3390/app9030372.

[70] B. Afzal, M. Umair, G. Asadullah Shah, E. Ahmed, Enabling IoT platforms for social IoT applications: Vision, feature mapping, and challenges, Futur. Gener. Comput. Syst. 92 (2019) 718–731. https://doi.org/10.1016/j.future.2017.12.002.

[71] M. Hosseinzadeh, V. Mohammadi, J. Lansky, V. Nulicek, Advancing the Social Internet of Things (SIoT): Challenges, Innovations, and Future Perspectives, Mathematics 12

[72] M. Becherer, O.K. Hussain, Y. Zhang, F. Den Hartog, E. Chang, On Trust Recommendations in the Social Internet of Things - A Survey, ACM Comput. Surv. 56 (2024). https://doi.org/10.1145/3645100.

[73] H.U. Khan, M. Abbas, F. Khan, S. Nazir, A. Binbusayyis, A. Alabdultif, W. Taegkeun, Multi-criteria decision-making methods for the evaluation of the social internet of things for the potential of defining human behaviors, Comput. Human Behav. 157 (2024). https://doi.org/10.1016/j.chb.2024.108230.

[74] F. Ullah, M.R. Naeem, L. Mostarda, S.A. Shah, Clone detection in 5G-enabled social IoT system using graph semantics and deep learning model, Int. J. Mach. Learn. Cybern. (2021). https://doi.org/10.1007/s13042-020-01246-9.

[75] A.E. Khaled, S. Helal, 2016 IEEE 3rd World Forum on Internet of Things, WF-IoT 2016, 2016 IEEE 3rd World Forum Internet Things, WF-IoT 2016 (2017).

[76] G.A. Stelea, V. Popescu, F. Sandu, L. Jalal, M. Farina, M. Murroni, From Things to Services: A Social IoT Approach for Tourist Service Management, IEEE Access 8 (2020) 153578–153588. https://doi.org/10.1109/ACCESS.2020.3018331.

[77] Z. Aghaee, A. Fatemi, P. Arebi, A controllability method on the social Internet of Things (SIoT) network, Pervasive Mob. Comput. 105 (2024). https://doi.org/10.1016/j.pmcj.2024.101992.

[78] R. Bi, Q. Chen, L. Chen, J. Xiong, D. Wu, A Privacy-Preserving Personalized Service Framework through Bayesian Game in Social IoT, Wirel. Commun. Mob. Comput. 2020 (2020). https://doi.org/10.1155/2020/8891889.

[79] C. Fu, C. Peng, X.Y. Liu, L.T. Yang, J. Yang, L. Han, Search engine: The social relationship driving power of Internet of Things, Futur. Gener. Comput. Syst. 92 (2019) 972–986. https://doi.org/10.1016/j.future.2018.01.037.

[80] O. Ben Abderrahim, M.H. Elhdhili, L. Saidane, TMCoI-SIOT: A trust management system based on communities of interest for the social Internet of Things, 2017 13th Int. Wirel. Commun. Mob. Comput. Conf. IWCMC 2017 (2017) 747–752. https://doi.org/10.1109/IWCMC.2017.7986378.

[81] M. Pateraki, K. Fysarakis, V. Sakkalis, G. Spanoudakis, I. Varlamis, M. Maniadakis, M. Lourakis, S. Ioannidis, N. Cummins, B. Schuller, E. Loutsetis, D. Koutsouris, Biosensors

(2024). https://doi.org/10.3390/math12050715.

and Internet of Things in smart healthcare applications: challenges and opportunities, in: Wearable Implant. Med. Devices, Elsevier, 2020: pp. 25–53. https://doi.org/10.1016/b978-0-12-815369-7.00002-1.

[82] R. Chataut, A. Phoummalayvane, R. Akl, Unleashing the Power of IoT: A Comprehensive Review of IoT Applications and Future Prospects in Healthcare, Agriculture, Smart Homes, Smart Cities, and Industry 4.0, Sensors 23 (2023). https://doi.org/10.3390/s23167194.

[83] N. Mrewa, A. Mohd Ramly, A. Amphawan, T.K. Neo, Optimizing Medical IoT Disaster Management with Data Compression, J. Informatics Web Eng. 3 (2024). https://doi.org/10.33093/jiwe.2024.3.1.4.

[84] J.F. Tesch, A.S. Brillinger, D. Bilgeri, Internet of things business model innovation and the stage-gate process: An exploratory analysis, Springer International Publishing, 2021. https://doi.org/10.1142/9781786347602_0008.

[85] F. Jindal, R. Jamar, P. Churi, Future and Challenges of Internet of Things, Int. J. Comput. Sci. Inf. Technol. 10 (2018) 13–25. https://doi.org/10.5121/ijcsit.2018.10202.

[86] X. Feng, J. Han, R. Zhang, S. Xu, H. Xia, Security defense strategy algorithm for Internet of Things based on deep reinforcement learning, High-Confidence Comput. 4 (2024). https://doi.org/10.1016/j.hcc.2023.100167.

[87] J.J. Zhang, K. Liu, F. Khalid, M.A. Hanif, S. Rehman, T. Theocharides, A. Artussi, M. Shafique, S. Garg, INVITED: Building robust machine learning systems: Current progress, research challenges, and opportunities, in: Proc. - Des. Autom. Conf., Institute of Electrical and Electronics Engineers Inc., 2019. https://doi.org/10.1145/3316781.3323472.

[88] K. Chopra, K. Gupta, A. Lambora, Future internet: The internet of things-a literature review, in: 2019 Int. Conf. Mach. Learn. Big Data, Cloud Parallel Comput., IEEE, 2019: pp. 135–139.

[89] F.P. Oikonomou, J. Ribeiro, G. Mantas, J.M.C.S. Bastos, J. Rodriguez, A hyperledger fabric-based blockchain architecture to secure iot-based health monitoring systems, in: 2021 IEEE Int. Mediterr. Conf. Commun. Netw., IEEE, 2021: pp. 186–190.

[90] K. Hameed, M. Barika, S. Garg, M.B. Amin, B. Kang, A taxonomy study on securing Blockchain-based Industrial applications: An overview, application perspectives, requirements, attacks, countermeasures, and open issues, J. Ind. Inf. Integr. 26 (2022)

100312.

[91] T. Janarthanan, M. Bagheri, S. Zargari, IoT forensics: An overview of the current issues and challenges, Digit. Forensic Investig. Internet Things Devices (2021) 223–254.

[92] A.A. Ahmed, K. Farhan, W.A. Jabbar, A. Al-Othmani, A.G. Abdulrahman, IoT Forensics: Current Perspectives and Future Directions, Sensors 24 (2024). https://doi.org/10.3390/s24165210.

[93] S. Munirathinam, Industry 4.0: Industrial Internet of Things (IIOT), Adv. Comput. 117 (2020) 129–164. https://doi.org/10.1016/BS.ADCOM.2019.10.010.

[94] D.H. Shin, S.J. Han, Y. Bin Kim, I.C. Euom, Research on Digital Forensics Analyzing Heterogeneous Internet of Things Incident Investigations, Appl. Sci. 14 (2024). https://doi.org/10.3390/app14031128.

[95] J.-P.A. Yaacoub, H.N. Noura, O. Salman, A. Chehab, Advanced digital forensics and anti-digital forensics for IoT systems: Techniques, limitations and recommendations, Internet of Things 19 (2022) 100544.

[96] M. Pathak, K.N. Mishra, S.P. Singh, Securing data and preserving privacy in cloud IoT-based technologies an analysis of assessing threats and developing effective safeguard, Artif. Intell. Rev. 57 (2024). https://doi.org/10.1007/s10462-024-10908-x.

[97] B.K. Sharma, M. Hachem, V.P. Mishra, M.J. Kaur, Internet of Things in forensics investigation in comparison to digital forensics, Handb. Wirel. Sens. Networks Issues Challenges Curr. Scenar. (2020) 672–684.

[98] S.S. Alqahtany, T.A. Syed, ForensicTransMonitor: A Comprehensive Blockchain Approach to Reinvent Digital Forensics and Evidence Management, Inf. 15 (2024). https://doi.org/10.3390/info15020109.

[99] J. Li, Z. Wang, S. Guan, Y. Cao, ProChain: A privacy-preserving blockchain-based supply chain traceability system model, Comput. Ind. Eng. 187 (2024). https://doi.org/10.1016/j.cie.2023.109831.

[100] M. Hossain, Y. Karim, R. Hasan, FIF-IoT: A forensic investigation framework for IoT using a public digital ledger, Proc. - 2018 IEEE Int. Congr. Internet Things, ICIOT 2018 - Part 2018 IEEE World Congr. Serv. (2018) 33–40. https://doi.org/10.1109/ICIOT.2018.00012.

[101] F. Bouchaud, G. Grimaud, T. Vantroys, IoT forensic: Identification and classification of

evidence in criminal investigations, ACM Int. Conf. Proceeding Ser. (2018). https://doi.org/10.1145/3230833.3233257.

[102] N.S.G. Ganesh, N.G.M. Venkatesh, D.V.V. Prasad, A systematic literature review on forensics in cloud, IoT, AI & blockchain, Illum. Artif. Intell. Cybersecurity Forensics (2022) 197–229.

[103] A.P. Renold, Survey of Evidence Collection Methods for Internet of Things Forensics, in: 2023 Int. Conf. Netw. Commun., IEEE, 2023: pp. 1–7.

[104] M.S. Mazhar, Y. Saleem, A. Almogren, J. Arshad, M.H. Jaffery, A.U. Rehman, M. Shafiq, H. Hamam, Forensic Analysis on Internet of Things ( IoT ) Device Using, (2022) 1–23.

[105] J. Kim, J. Park, S. Lee, An improved IoT forensic model to identify interconnectivity between things, Forensic Sci. Int. Digit. Investig. 44 (2023). https://doi.org/10.1016/j.fsidi.2022.301499.

[106] V.R. Kebande, N.M. Karie, H.S. Venter, Cloud-Centric Framework for isolating Big data as forensic evidence from IoT infrastructures, 2017 1st Int. Conf. Next Gener. Comput. Appl. NextComp 2017 (2017) 54–60. https://doi.org/10.1109/NEXTCOMP.2017.8016176.

[107] L. Sana, M.M. Nazir, J. Yang, L. Hussain, Y.L. Chen, C.S. Ku, M. Alatiyyah, S.A. Alateyah, L.Y. Por, Securing the IoT Cyber Environment: Enhancing Intrusion Anomaly Detection With Vision Transformers, IEEE Access 12 (2024) 82443–82468. https://doi.org/10.1109/ACCESS.2024.3404778.

[108] P. Singh, C. Engineering, Cross-Layer Design for Internet of Things ( IOT ) -Issues and Possible Solutions, Dep. Syst. Comput. Eng. (2018) 1–10.

[109] F.J. Egloff, M. Smeets, Publicly attributing cyber attacks: a framework, J. Strateg. Stud. (2021) 1–32.

[110] E. Schiller, A. Aidoo, J. Fuhrer, J. Stahl, M. Ziörjen, B. Stiller, Landscape of IoT security, Comput. Sci. Rev. 44 (2022) 100467.

[111] P. Malhotra, Y. Singh, P. Anand, D.K. Bangotra, P.K. Singh, W.-C. Hong, Internet of things: Evolution, concerns and security challenges, Sensors 21 (2021) 1809.

[112] J.P.A. Yaacoub, H.N. Noura, O. Salman, A. Chehab, Ethical hacking for IoT: Security issues, challenges, solutions and recommendations, Internet Things Cyber-Physical Syst. 3

(2023). https://doi.org/10.1016/j.iotcps.2023.04.002.

[113] O. Sadio, I. Ngom, C. Lishou, Lightweight security scheme for mqtt/mqtt-sn protocol, in: 2019 Sixth Int. Conf. Internet Things Syst. Manag. Secur., IEEE, 2019: pp. 119–123.

[114] N.L. Fantana, T. Riedel, J. Schlick, S. Ferber, J. Hupp, S. Miles, F. Michahelles, S. Svensson, IoT Applications — Value Creation for Industry, in: Internet of Things, 2022. https://doi.org/10.1201/9781003338659-3.

[115] A. Hameed, A. Alomary, Security issues in IoT: A survey, 2019 Int. Conf. Innov. Intell. Informatics, Comput. Technol. 3ICT 2019 (2019) 1–5. https://doi.org/10.1109/3ICT.2019.8910320.

[116] M. Abdur Razzaq, S. Abdul Latif University Khairpur, P. Sajid Habib Gill, E. Rahim Yar Khan Campus Rahim Yar Khan, P. Muhammad Ali Qureshi, S. Ullah, R. Yar Khan, Security Issues in the Internet of Things (IoT): A Comprehensive Study, IJACSA) Int. J. Adv. Comput. Sci. Appl. 8 (2017).

[117] U. Chatterjee, S. Ray, Security Issues on IoT Communication and Evolving Solutions, Stud. Comput. Intell. 988 (2022) 183–204. https://doi.org/10.1007/978-981-16-4713-0_10.

[118] N. Siddiqui, L. Pryor, R. Dave, User authentication schemes using machine learning methods—a review, in: Proc. Int. Conf. Commun. Comput. Technol. ICCCT 2021, Springer, 2021: pp. 703–723.

[119] M. Kokila, S. Reddy K, Authentication, access control and scalability models in Internet of Things Security–A review, Cyber Secur. Appl. 3 (2025). https://doi.org/10.1016/j.csa.2024.100057.

[120] M. El-Hajj, A. Fadlallah, M. Chamoun, A. Serhrouchni, A Survey of Internet of Things (IoT) Authentication Schemes, Sensors 2019, Vol. 19, Page 1141 19 (2019) 1141. https://doi.org/10.3390/S19051141.

[121] A. Munir, I.A. Sumra, R. Naveed, M.A. Javed, Techniques for Authentication and Defense Strategies to Mitigate IoT Security Risks, J. Comput. Biomed. Informatics 07 (2024) 1–12.

[122] A. ur Rehman, S. ur Rehman, I.U. Khan, M. Moiz, S. Hasan, Security and Privacy Issues in IoT, Int. J. Commun. Networks Inf. Secur. 8 (2016). https://doi.org/10.17762/IJCNIS.V8I3.2074.

[123] T. Nandy, M.Y.I. Bin Idris, R. Md Noor, M.L. Mat Kiah, L.S. Lun, N.B. Annuar Juma'At, I. Ahmedy, N. Abdul Ghani, S. Bhattacharyya, Review on Security of Internet of Things Authentication Mechanism, IEEE Access 7 (2019) 151054–151089. https://doi.org/10.1109/ACCESS.2019.2947723.

[124] D. Li, W. Peng, W. Deng, F. Gai, A blockchain-based authentication and security mechanism for IoT, Proc. - Int. Conf. Comput. Commun. Networks, ICCCN 2018-July (2018). https://doi.org/10.1109/ICCCN.2018.8487449.

[125] A. N.alsheavi, A. Hawbani, X. Wang, W. Othman, L. Zhao, Z. Liu, S.H. Alsamhi, M.A.A. Al-Qaness, IoT Authentication Protocols: Classification, Trend and Opportunities, IEEE Trans. Sustain. Comput. (2024). https://doi.org/10.1109/TSUSC.2024.3492152.

[126] S. Hameed, F.I. Khan, B. Hameed, Understanding Security Requirements and Challenges in Internet of Things (IoT): A Review, J. Comput. Networks Commun. 2019 (2019). https://doi.org/10.1155/2019/9629381.

[127] A. Ometov, S. Bezzateev, N. Mäkitalo, S. Andreev, T. Mikkonen, Y. Koucheryavy, Multi-factor authentication: A survey, Cryptography 2 (2018) 1–31. https://doi.org/10.3390/CRYPTOGRAPHY2010001.

[128] M. Tanveer, S.A. Aldossari, RAM-MEN: Robust authentication mechanism for IoT-enabled edge networks, Alexandria Eng. J. 112 (2025) 436–447. https://doi.org/10.1016/j.aej.2024.10.116.

[129] T. Shah, S. Venkatesan, Authentication of IoT Device and IoT Server Using Secure Vaults, Proc. - 17th IEEE Int. Conf. Trust. Secur. Priv. Comput. Commun. 12th IEEE Int. Conf. Big Data Sci. Eng. Trust. 2018 (2018) 819–824. https://doi.org/10.1109/TRUSTCOM/BIGDATASE.2018.00117.

[130] S. Biswas, K. Sharif, F. Li, B. Nour, Y. Wang, A scalable blockchain framework for secure transactions in IoT, IEEE Internet Things J. 6 (2019) 4650–4659. https://doi.org/10.1109/JIOT.2018.2874095.

[131] J. Huang, L. Kong, G. Chen, M.Y. Wu, X. Liu, P. Zeng, Towards secure industrial iot: Blockchain system with credit-based consensus mechanism, IEEE Trans. Ind. Informatics 15 (2019) 3680–3689. https://doi.org/10.1109/TII.2019.2903342.

[132] M.T. Hammi, B. Hammi, P. Bellot, A. Serhrouchni, Bubbles of Trust: A decentralized blockchain-based authentication system for IoT, Comput. Secur. 78 (2018) 126–142.

https://doi.org/10.1016/J.COSE.2018.06.004.

[133] R. Anitha, B.R. Tapas Bapu, Blockchain-based Light-weight Authentication Approach for a Multiple Wireless Sensor Network, IETE J. Res. 70 (2024). https://doi.org/10.1080/03772063.2022.2154710.

[134] U. Khalid, M. Asim, T. Baker, P.C.K. Hung, M.A. Tariq, L. Rafferty, A decentralized lightweight blockchain-based authentication mechanism for IoT systems, Cluster Comput. 23 (2020) 2067–2087. https://doi.org/10.1007/S10586-020-03058-6/FIGURES/11.

[135] A.K. Das, A secure and robust temporal credential-based three-factor user authentication scheme for wireless sensor networks, Peer-to-Peer Netw. Appl. 9 (2016) 223–244. https://doi.org/10.1007/S12083-014-0324-9/TABLES/7.

[136] H. Nicanfar, P. Jokar, K. Beznosov, V.C.M. Leung, Efficient authentication and key management mechanisms for smart grid communications, IEEE Syst. J. 8 (2014) 629–640. https://doi.org/10.1109/JSYST.2013.2260942.

[137] Q. Jiang, N. Kumar, J. Ma, J. Shen, D. He, N. Chilamkurti, A privacy-aware two-factor authentication protocol based on elliptic curve cryptography for wireless sensor networks, Int. J. Netw. Manag. 27 (2017) e1937. https://doi.org/10.1002/NEM.1937.

[138] C.H. Lee, K.H. Kim, Implementation of IoT system using block chain with authentication and data protection, Int. Conf. Inf. Netw. 2018-Janua (2018) 936–940. https://doi.org/10.1109/ICOIN.2018.8343261.

[139] G.M. Abdullah, Q. Mehmood, C.B.A. Khan, Adoption of Lamport signature scheme to implement digital signatures in IoT, 2018 Int. Conf. Comput. Math. Eng. Technol. Inven. Innov. Integr. Socioecon. Dev. ICoMET 2018 - Proc. 2018-Janua (2018) 1–4. https://doi.org/10.1109/ICOMET.2018.8346359.

[140] S. Ravidas, A. Lekidis, F. Paci, N. Zannone, Access control in Internet-of-Things: A survey, J. Netw. Comput. Appl. 144 (2019) 79–101. https://doi.org/10.1016/j.jnca.2019.06.017.

[141] Y. Zhang, S. Kasahara, Y. Shen, X. Jiang, J. Wan, Smart contract-based access control for the internet of things, IEEE Internet Things J. 6 (2018) 1594–1605.

[142] C.-L. Hsu, W.-X. Chen, T.-V. Le, An Autonomous Log Storage Management Protocol with Blockchain Mechanism and Access Control for the Internet of Things, Sensors 20 (2020) 6471. https://doi.org/10.3390/s20226471.

[143] R.Y. Patil, A secure privacy preserving and access control scheme for medical internet of things (MIoT) using attribute-based signcryption, Int. J. Inf. Technol. 16 (2024). https://doi.org/10.1007/s41870-023-01569-0.

[144] S. Pal, M. Hitchens, V. Varadharajan, T. Rabehaja, Policy-based access control for constrained healthcare resources in the context of the Internet of Things, J. Netw. Comput. Appl. 139 (2019) 57–74. https://doi.org/10.1016/j.jnca.2019.04.013.

[145] L. Sadineni, E. Pilli, R.B. Battula, A Holistic Forensic Model for the Internet of Things, in: 2019: pp. 3–18. https://doi.org/10.1007/978-3-030-28752-8_1.

[146] M. Ammar, G. Russello, B. Crispo, Internet of Things: A survey on the security of IoT frameworks, J. Inf. Secur. Appl. 38 (2018) 8–27. https://doi.org/10.1016/J.JISA.2017.11.002.

[147] T. Yu, V. Sekar, S. Seshan, Y. Agarwal, C. Xu, Handling a trillion (unfixable) flaws on a billion devices: Rethinking network security for the Internet-of-Things, Proc. 14th ACM Work. Hot Top. Networks, HotNets-XIV 2015 (2015). https://doi.org/10.1145/2834050.2834095.

[148] M. Seliem, K. Elgazzar, K. Khalil, Towards Privacy Preserving IoT Environments: A Survey, Wirel. Commun. Mob. Comput. 2018 (2018). https://doi.org/10.1155/2018/1032761.

[149] R. Vignesh, A. Samydurai, Security on Internet of Things (IOT) with Challenges and Countermeasures, J. Sci. Res. 12 (2017) 128–134. https://doi.org/10.5829/idosi.aejsr.2017.128.134.

[150] D. Miorandi, S. Sicari, F. De Pellegrini, I. Chlamtac, Internet of things: Vision, applications and research challenges, Ad Hoc Networks 10 (2012) 1497–1516. https://doi.org/10.1016/J.ADHOC.2012.02.016.

[151] M. Usman, I. Ahmed, M.I. Aslam, S. Khan, U.A. Shah, SIT: A Lightweight Encryption Algorithm for Secure Internet of Things, Int. J. Adv. Comput. Sci. Appl. 8 (2017) 402–411. https://doi.org/10.14569/IJACSA.2017.080151.

[152] S. Bansal, D. Kumar, IoT ecosystem: A survey on devices, gateways, operating systems, middleware and communication, Int. J. Wirel. Inf. Networks 27 (2020) 340–364.

[153] A. Ghaffari, N. Jelodari, S. pouralish, N. derakhshanfard, B. Arasteh, Securing internet of things using machine and deep learning methods: a survey, Cluster Comput. 27 (2024)

9065–9089. https://doi.org/10.1007/s10586-024-04509-0.

[154] U.K. Lilhore, S. Dalal, S. Simaiya, A cognitive security framework for detecting intrusions in IoT and 5G utilizing deep learning, Comput. Secur. 136 (2024). https://doi.org/10.1016/j.cose.2023.103560.

[155] N. Koroniotis, N. Moustafa, E. Sitnikova, A new network forensic framework based on deep learning for Internet of Things networks: A particle deep framework, Futur. Gener. Comput. Syst. 110 (2020) 91–106. https://doi.org/10.1016/j.future.2020.03.042.

[156] Hemani, D. Singh, R.K. Dwivedi, Designing blockchain based secure autonomous vehicular internet of things (IoT) architecture with efficient smart contracts, Int. J. Inf. Technol. (2024). https://doi.org/10.1007/s41870-023-01712-x.

[157] P.M. Shakeel, S. Baskar, H. Fouad, G. Manogaran, V. Saravanan, C.E. Montenegro-Marin, Internet of things forensic data analysis using machine learning to identify roots of data scavenging, Futur. Gener. Comput. Syst. 115 (2021) 756–768. https://doi.org/10.1016/j.future.2020.10.001.

[158] B.K. Mohanta, D. Jena, U. Satapathy, S. Patnaik, Survey on IoT security: Challenges and solution using machine learning, artificial intelligence and blockchain technology, Internet of Things 11 (2020) 100227.

[159] Z.A. Baig, S. Sanguanpong, S.N. Firdous, V.N. Vo, T.G. Nguyen, C. So-In, Averaged dependence estimators for DoS attack detection in IoT networks, Futur. Gener. Comput. Syst. 102 (2020) 198–209. https://doi.org/https://doi.org/10.1016/j.future.2019.08.007.

[160] D. Yin, L. Zhang, K. Yang, A DDoS attack detection and mitigation with software-defined Internet of Things framework, IEEE Access 6 (2018) 24694–24705.

[161] M. López, A. Peinado, A. Ortiz, An extensive validation of a SIR epidemic model to study the propagation of jamming attacks against IoT wireless networks, Comput. Networks 165 (2019) 106945.

[162] M. Almiani, A. AbuGhazleh, A. Al-Rahayfeh, S. Atiewi, A. Razaque, Deep recurrent neural network for IoT intrusion detection system, Simul. Model. Pract. Theory 101 (2020) 102031.

[163] I. Ali, A.I.A. Ahmed, A. Almogren, M.A. Raza, S.A. Shah, A. Khan, A. Gani, Systematic literature review on IoT-based botnet attack, IEEE Access 8 (2020) 212220–212232.

[164] L. Liu, Z. Ma, W. Meng, Detection of multiple-mix-attack malicious nodes using

perceptron-based trust in IoT networks, Futur. Gener. Comput. Syst. 101 (2019) 865–879. https://doi.org/https://doi.org/10.1016/j.future.2019.07.021.

[165] J. Moon, I.Y. Jung, J.H. Park, IoT application protection against power analysis attack, Comput. Electr. Eng. 67 (2018) 566–578. https://doi.org/https://doi.org/10.1016/j.compeleceng.2018.02.030.

[166] G. Rajendran, R.S.R. Nivash, P.P. Parthy, S. Balamurugan, Modern security threats in the Internet of Things (IoT): Attacks and Countermeasures, in: 2019 Int. Carnahan Conf. Secur. Technol., IEEE, 2019: pp. 1–6.

[167] C. Li, Z. Qin, E. Novak, Q. Li, Securing SDN infrastructure of IoT–fog networks from MitM attacks, IEEE Internet Things J. 4 (2017) 1156–1164.

[168] J. Deogirikar, A. Vidhate, Security attacks in IoT: A survey, Proc. Int. Conf. IoT Soc. Mobile, Anal. Cloud, I-SMAC 2017 (2017) 32–37. https://doi.org/10.1109/I-SMAC.2017.8058363.

[169] S. Deshmukh-Bhosale, S.S. Sonavane, A Real-Time Intrusion Detection System for Wormhole Attack in the RPL based Internet of Things, Procedia Manuf. 32 (2019) 840–847. https://doi.org/https://doi.org/10.1016/j.promfg.2019.02.292.

[170] S.E. Ali, N. Tariq, F.A. Khan, M. Ashraf, W. Abdul, K. Saleem, BFT-IoMT: A Blockchain-Based Trust Mechanism to Mitigate Sybil Attack Using Fuzzy Logic in the Internet of Medical Things, Sensors 23 (2023) 4265.

[171] A. Blaise, M. Bouet, V. Conan, S. Secci, Detection of zero-day attacks: An unsupervised port-based approach, Comput. Networks 180 (2020) 107391.

[172] B. Xu, W. Wang, Q. Hao, Z. Zhang, P. Du, T. Xia, H. Li, X. Wang, A security design for the detecting of buffer overflow attacks in IoT device, IEEE Access 6 (2018) 72862–72869.

[173] S. Eggers, A novel approach for analyzing the nuclear supply chain cyber-attack surface, Nucl. Eng. Technol. 53 (2021) 879–887.

[174] S. Tu, M. Waqas, S.U. Rehman, M. Aamir, O.U. Rehman, Z. Jianbiao, C.-C. Chang, Security in fog computing: A novel technique to tackle an impersonation attack, IEEE Access 6 (2018) 74993–75001.

[175] J.P.A. Yaacoub, M. Noura, H.N. Noura, O. Salman, E. Yaacoub, R. Couturier, A. Chehab, Securing internet of medical things systems: Limitations, issues and recommendations,

Futur. Gener. Comput. Syst. 105 (2020). https://doi.org/10.1016/j.future.2019.12.028.

[176] I. Nadir, H. Mahmood, G. Asadullah, A taxonomy of IoT firmware security and principal firmware analysis techniques, Int. J. Crit. Infrastruct. Prot. (2022) 100552.

[177] A. Javed, K.M. Malik, A. Irtaza, H. Malik, Towards protecting cyber-physical and IoT systems from single-and multi-order voice spoofing attacks, Appl. Acoust. 183 (2021) 108283.

[178] P.J. Ryan, R.B. Watson, Research Challenges for the Internet of Things: What Role Can OR Play?, Syst. 2017, Vol. 5, Page 24 5 (2017) 24. https://doi.org/10.3390/SYSTEMS5010024.

[179] A. Abdelmaboud, A. Ibrahim Abdalla Ahmed, M. Abaker, T. Abdalla Elfadil Eisa, H. Albasheer, S. Abdelwahab Ghorashi, F. Khalid Karim, Blockchain for IoT Applications: Taxonomy, Platforms, Recent Advances, Challenges and Future Research Directions, Electron. 2022, Vol. 11, Page 630 11 (2022) 630. https://doi.org/10.3390/ELECTRONICS11040630.

[180] T. Mazhar, S.F.A. Shah, S.A. Inam, J.B. Awotunde, M.M. Saeed, H. Hamam, Analysis of integration of IoMT with blockchain: issues, challenges and solutions, Discov. Internet Things 4 (2024). https://doi.org/10.1007/s43926-024-00078-1.

[181] M.A. Uddin, A. Stranieri, I. Gondal, V. Balasubramanian, A survey on the adoption of blockchain in IoT: challenges and solutions, Blockchain Res. Appl. 2 (2021) 100006. https://doi.org/10.1016/J.BCRA.2021.100006.

[182] M. Andoni, V. Robu, D. Flynn, S. Abram, D. Geach, D. Jenkins, P. McCallum, A. Peacock, Blockchain technology in the energy sector: A systematic review of challenges and opportunities, Renew. Sustain. Energy Rev. 100 (2019) 143–174. https://doi.org/10.1016/J.RSER.2018.10.014.

[183] A.S. Raikar, P. Kumar, G. (Vedant) S. Raikar, S.N. Somnache, Advances and Challenges in IoT-Based Smart Drug Delivery Systems: A Comprehensive Review, Appl. Syst. Innov. 6 (2023). https://doi.org/10.3390/asi6040062.

[184] Kudirat Bukola Adeusi, Ayodeji Enoch Adegbola, Prisca Amajuoyi, Mayokun Daniel Adegbola, Lucky Bamidele Benjamin, The potential of IoT to transform supply chain management through enhanced connectivity and real-time data, World J. Adv. Eng. Technol. Sci. 12 (2024) 145–151. https://doi.org/10.30574/wjaets.2024.12.1.0202.

[185] E. Jonas, J. Schleier-Smith, V. Sreekanti, C.-C. Tsai, A. Khandelwal, Q. Pu, V. Shankar, J. Carreira, K. Krauth, N. Yadwadkar, J.E. Gonzalez, R.A. Popa, I. Stoica, D.A. Patterson, Cloud Programming Simplified: A Berkeley View on Serverless Computing, (2019).

[186] A.J. Onumanyi, A.M. Abu-Mahfouz, G.P. Hancke, Low Power Wide Area Network, Cognitive Radio and the Internet of Things: Potentials for Integration, Sensors 2020, Vol. 20, Page 6837 20 (2020) 6837. https://doi.org/10.3390/S20236837.

[187] A. Srivastava, S. Gupta, M. Quamara, P. Chaudhary, V.J. Aski, Future IoT-enabled threats and vulnerabilities: State of the art, challenges, and future prospects, Int. J. Commun. Syst. 33 (2020) e4443. https://doi.org/10.1002/DAC.4443.

[188] Q. Zhao, Z. Peng, X. Hong, A Named Data Networking Architecture Implementation to Internet of Underwater Things, (2019). https://doi.org/10.1145/3366486.3366506.

[189] R.K. Dudeja, R. Singh Bali, G.S. Aujla, An Efficient Data Communication Framework for Named Data Networking in IoT-Edge Ecosystem, (2022) 1–6. https://doi.org/10.1109/GCWKSHPS52748.2021.9681976.

[190] N.N. Thilakarathne, Security and Privacy Issues in IoT Environment, Int. J. Eng. Manag. Res. 10 (2020) 26–29. https://doi.org/10.31033/IJEMR.10.1.5.

[191] J. Chaudhry, K. Saleem, P. Haskell-Dowland, M.H. Miraz, A Survey of Distributed Certificate Authorities in MANETs, Ann. Emerg. Technol. Comput. 2 (2018) 11–18. https://doi.org/10.33166/AETiC.2018.03.002.

[192] A.S. Abu Daia, R.A. Ramadan, M.B. Fayek, Sensor networks attacks classifications and mitigation, Ann. Emerg. Technol. Comput. 2 (2018) 28–43. https://doi.org/10.33166/AETIC.2018.04.003.

[193] M.H. Miraz, M. Ali, P.S. Excell, R. Picking, Internet of Nano-Things, things and everything: Future growth trends, Futur. Internet 10 (2018). https://doi.org/10.3390/FI10080068.

[194] M.H. Miaz, M. Ali, Integration of blockchain and IoT: An enhanced security perspective, Ann. Emerg. Technol. Comput. 4 (2020) 52–63. https://doi.org/10.33166/AETIC.2020.04.006.

[195] I. Andrea, C. Chrysostomou, G. Hadjichristofi, Internet of Things: Security vulnerabilities and challenges, Proc. - IEEE Symp. Comput. Commun. 2016-Febru (2016) 180–187. https://doi.org/10.1109/ISCC.2015.7405513.

[196] N.N. Srinidhi, S.M. Dilip Kumar, K.R. Venugopal, Network optimizations in the Internet of Things: A review, Eng. Sci. Technol. an Int. J. 22 (2019) 1–21. https://doi.org/10.1016/J.JESTCH.2018.09.003.

[197] J. Pacheco, S. Hariri, IoT security framework for smart cyber infrastructures, Proc. - IEEE 1st Int. Work. Found. Appl. Self-Systems, FAS-W 2016 (2016) 242–247. https://doi.org/10.1109/FAS-W.2016.58.

[198] S. Al Salami, J. Baek, K. Salah, E. Damiani, Lightweight encryption for smart home, Proc. - 2016 11th Int. Conf. Availability, Reliab. Secur. ARES 2016 (2016) 382–388. https://doi.org/10.1109/ARES.2016.40.

[199] A. Al-Omary, H.M. AlSabbagh, H. Al-Rizzo, Survey of Hardware-based Security support for IoT/CPS Systems, KnE Eng. 3 (2018) 52. https://doi.org/10.18502/KEG.V3I7.3072.

[200] H. Shin, H.K. Lee, H.Y. Cha, S.W. Heo, H. Kim, IoT Security Issues and Light Weight Block Cipher, 1st Int. Conf. Artif. Intell. Inf. Commun. ICAIIC 2019 (2019) 381–384. https://doi.org/10.1109/ICAIIC.2019.8669029.

[201] S. Upadhyay, M. Kumar, A. Upadhyay, S. Verma, Kavita, M. Kaur, R.A. Khurma, P.A. Castillo, Challenges and Limitation Analysis of an IoT-Dependent System for Deployment in Smart Healthcare Using Communication Standards Features, Sensors 23 (2023). https://doi.org/10.3390/s23115155.

[202] A. Rekeraho, D.T. Cotfas, P.A. Cotfas, T.C. Bălan, E. Tuyishime, R. Acheampong, Cybersecurity challenges in IoT-based smart renewable energy, Int. J. Inf. Secur. 23 (2024). https://doi.org/10.1007/s10207-023-00732-9.

[203] S.A. Al-Qaseemi, H.A. Almulhim, M.F. Almulhim, S.R. Chaudhry, IoT architecture challenges and issues: Lack of standardization, FTC 2016 - Proc. Futur. Technol. Conf. (2017) 731–738. https://doi.org/10.1109/FTC.2016.7821686.

[204] A.A. Ahmed, M.M. AL-Shaboti, Implementation of Internet of Things (IoT) Based on IPv6 over Wireless Sensor Networks, Int. J. Sensors, Wirel. Commun. Control 7 (2017). https://doi.org/10.2174/2210327907666170911145726.

[205] H.U. Rehman, M. Asif, M. Ahmad, Future applications and research challenges of IOT, 2017 Int. Conf. Inf. Commun. Technol. ICICT 2017 2017-Decem (2018) 68–74. https://doi.org/10.1109/ICICT.2017.8320166.

[206] A. Čolaković, M. Hadžialić, Internet of Things (IoT): A review of enabling technologies,

challenges, and open research issues, Comput. Networks 144 (2018) 17–39. https://doi.org/10.1016/J.COMNET.2018.07.017.

[207] A. Pal, H.K. Rath, S. Shailendra, AbhijanBhattacharyya, IoT Standardization: The Road Ahead, Internet Things - Technol. Appl. Stand. (2018). https://doi.org/10.5772/INTECHOPEN.75137.

[208] S.K. Datta, C. Bonnet, Integrating Named Data Networking in Internet of Things architecture, 2016 IEEE Int. Conf. Consum. Electron. ICCE-TW 2016 (2016). https://doi.org/10.1109/ICCE-TW.2016.7520967.

[209] B. Guo, D. Zhang, Z. Wang, Z. Yu, X. Zhou, Opportunistic IoT: Exploring the harmonious interaction between human and the internet of things, J. Netw. Comput. Appl. 36 (2013) 1531–1539. https://doi.org/10.1016/J.JNCA.2012.12.028.

[210] A. Rao, O. Schelén, A. Lindgren, Performance implications for IoT over information centric networks, Proc. Annu. Int. Conf. Mob. Comput. Networking, MOBICOM (2016) 57–62. https://doi.org/10.1145/2979683.2979686.

[211] F.M. Al-Turjman, A.E. Al-Fagih, W.M. Alsalih, H.S. Hassanein, A delay-tolerant framework for integrated RSNs in IoT, Comput. Commun. 36 (2013) 998–1010. https://doi.org/10.1016/J.COMCOM.2012.07.001.

[212] M.C. Domingo, An overview of the internet of underwater things, J. Netw. Comput. Appl. 35 (2012) 1879–1890. https://doi.org/10.1016/J.JNCA.2012.07.012.

[213] B. Chatterjee, S. Sen, N. Cao, A. Raychowdhury, Context-Aware Intelligence in Resource-Constrained IoT Nodes: Opportunities and Challenges, IEEE Des. Test 36 (2019) 7–40. https://doi.org/10.1109/MDAT.2019.2899334.